**Paleo-Rock-Hosted Life on Earth and the Search on Mars: a Review and Strategy for Exploration**

T.C. Onstott[1,*], B.L. Ehlmann[2,3,*], H. Sapers[2,3,4], M. Coleman[3,5], M. Ivarsson[6], J.J. Marlow[7], A. Neubeck[8], and P. Niles[9]

[1] Department of Geosciences, Princeton University, Princeton, NJ 008544 USA
[2] Division of Geological & Planetary Sciences, California Institute of Technology, Pasadena, CA 91125 USA
[3] Jet Propulsion Laboratory, California Institute of Technology, Pasadena, CA 91109 USA
[4] Department of Earth Sciences, University of Southern California, Los Angeles, CA 90089 USA
[5] NASA Astrobiology Institute, Pasadena California CA 91109 USA
[6] Department of Biology, University of Southern Denmark, DK-5230, Odense, Denmark
[7] Department of Organismic & Evolutionary Biology, Harvard University, Cambridge, MA 02138 USA
[8] Department of Earth Sciences, Upsala University, 752 36 Uppsala, Sweden
[9] Astromaterials Research and Exploration Science Division, NASA Johnson Space Center, Houston, TX 77058 USA

*These authors contributed equally to this work. Addresses for correspondence
tullis@princeton.edu, ehlmann@caltech.edu

**Abstract**

Here we review published studies on the abundance and diversity of terrestrial rock-hosted life, the environments it inhabits, the evolution of its metabolisms, and its fossil biomarkers to provide guidance in the search for the biomarkers of rock-hosted life on Mars. Key findings are: 1) the metabolic pathways for chemolithoautotrophic microorganisms evolved much earlier in Earth's history than those of surface-dwelling phototrophic microorganisms; 2) the emergence of the former occurred at a time when Mars was habitable, whereas the emergence of the latter occurred at a time when the martian surface would have been uninhabitable; 3) subsurface microbial cell concentrations do not correlate with the abundance of organic carbon and tend to be highest at interfaces where chemical redox gradients are most pronounced; 4) deep subsurface metabolic activity does not rely upon the respiration of organic photosynthate but upon the flux of inorganic energy and the abiotic and biotic recycling of metabolic waste products; 5) the rock record reveals examples of biomarkers of subsurface life at least back to 3.45 Ga with several examples of good preservation potential in rock types that are quite different from those preserving the photospheric-supported biosphere. These findings suggest that rock-hosted life would have been both more likely to emerge and be preserved in a martian context. With this in mind, we propose a Mars exploration strategy for subsurface life that scales spatially, focusing initially on identifying rocks with evidence for groundwater flow and low-temperature mineralization, then identifying redox and permeability interfaces preserved within rock outcrops, and finally focusing on finding minerals associated with redox reactions and traces of carbon and diagnostic biosignatures. The lessons from Earth show that ancient rock-hosted life is preserved in the fossil record and confirmable via a suite of morphologic, organic, mineralogical and isotopic fingerprints and micrometer-scale textures.

**1. Introduction**

From the mid-1980's to early 1990's evidence accumulated from both the continental and marine realms of a vast, well-populated underground biosphere that was on par with the total biomass on the Earth's surface (Onstott, 2016). The discovery by Stevens and McKinley (Stevens and McKinley, 1995) of subsurface lithoautotrophic microbial ecosystems (SLiMEs) in basaltic aquifers fueled by $H_2$ that was generated by reaction of water with Fe-bearing minerals had an immediate impact on the planetary science community, especially with respect to the search of extant life on Mars (McKay, 2001). Subsequently, subsurface life on Earth has been discovered at depths of 4-5 km in the continental crust (Moser *et al.*, 2005) and 2.5 km in sub-seafloor sediments (Inagaki *et al.*, 2015), at temperatures from -54 to 129°C, at pH values ranging from 3 to 13, and in solutions with ionic strengths up to 7M for continental crust sites (Magnabosco *et al.*, 2018a). Subsurface life is pervasive on Earth and rock-based microenvironments offer physical and energetic advantages to their inhabitants compared to the oceans and surface photosphere. In this paper we refer to "rock-hosted" life, whose existence is critically dependent upon physicochemical processes within the host rock, e.g., water-mineral, gaseous, or radiolytic reactions.

The most recent estimate of the mass of the Earth's subsurface biosphere is ~$10^{30}$ cells, which is about 10% of the surface biosphere (Magnabosco *et al.*, 2018a). One key question when considering the likelihood of finding subsurface life on other planets is how the abundance of Earth's subsurface life may have changed with time, coupled with the evolution and proliferation of surface life, e.g., beginning with the extensive colonization of land by plant life ~450 million years ago. Some portion of Earth's current global subsurface biosphere is supported directly or indirectly (through thermogenesis) by organic photosynthate from the surface biosphere while another portion is supported by abiotically-produced organic matter or autotrophic carbon fixation. In this paper we are careful to draw the distinctions between these two types of subsurface ecosystems. Over the last decade, it has become apparent that, rather than subsurface microbial communities originating as dying vestiges of transported surface microorganisms, the identification of novel species with no known closely related homologues is evidence for specialized, endemic communities of rock-hosted life (Chivian *et al.*, 2008; Momper *et al.*, 2017; Osburn *et al.*, 2014). In deep crustal environments rock-hosted life has been found to comprise entire ecosystems with multiple trophic levels (Lau *et al.*, 2016).

Habitats for rock-hosted life may have been—and may still be—present elsewhere in the solar system (Fig. 1). The sub-ocean silicate crusts of Europa and Enceladus have been proposed to host low-temperature groundwater/hydrothermal systems, leading to chemical/radiolytic reactions, which could supply energy for life (Deamer and Damer, 2017; Hand *et al.*, 2007; McKay *et al.*, 2012; McKay *et al.*, 2008; Pasek and Greenberg, 2012; Schluze-Makuch and Irwin, 2002; Steel *et al.*, 2017; Vance *et al.*, 2016). On Mars, throughout the first 1.5 billion years of its history, surface waters were intermittently present (Fassett and Head, 2011), whereas a more persistent and volumetrically more extensive aqueous environment existed beneath the surface, hosted in crystalline and sedimentary rocks (Clifford *et al.*, 2010; Clifford and Parker, 2001; Cockell, 2014a; Cockell, 2014b; Des Marais, 2010; Ehlmann *et al.*, 2011) and subsurface brines may still exist today (Orosei *et al.*, 2018). Because of the relative hostility and instability of the martian surface environment—aridity, sub-freezing temperatures, frequent climate change due to obliquity cycles, and radiation—compared to the comparatively clement and stable subsurface, sampling rock units that have or may have hosted groundwater warrants top priority in the search for life on Mars.

[Figure 1: Graphic of Rock-Hosted Habitats on Earth and Planets]

In this review we focus on the search for past rock-hosted life on Mars drawing on the lessons from Earth's record of extant and fossil rock-hosted life. We first describe the environmental history and habitability of Mars. We then review what is currently known about the extent, metabolic diversity, and community structure of present rock-hosted life on Earth, as well as its metabolic products. We next examine the evolutionary history of the enzymes utilized by rock-hosted versus photosynthetic life. We then address how long Earth's rock-hosted life communities, as evident in their biomarkers, have existed and what processes promote preservation of their morphological, mineralogical, isotopic and chemical traces in the rock record. Finally, we consider the large volumes of rock that constitute past and present habitable environments on Mars and articulate an operational strategy for their exploration for the biosignatures of rock-hosted life.

## 2. The Case For Targeting The Search For Life On Mars To Rock-Hosted Life

On Earth, extensive plate tectonics-driven crustal recycling has removed much of the earliest geologic record and metamorphosed the rest, obscuring the history and extent of the biosphere. On Mars, the ancient geologic record remains largely in place with >50% of the martian record at the surface preserved from earlier than 3.5 billion years (e.g. (Tanaka *et al.*, 2014), including ancient units uncovered by tectonics, erosion, and impact cratering. As such, if life evolved on Mars contemporaneously with Earth's life, the rocks and biosignatures recording the trajectory of its early evolution are better preserved and more easily accessible than those of time-equivalent periods on Earth.

Over the last decade, *in situ* exploration by rovers and high-resolution mineralogy and stratigraphy by orbiting instruments have revealed the nature of environmental conditions during the first ~1.5 billion years. Globally widespread phyllosilicate minerals (smectites, chlorites, and other hydrated silicates) were formed by aqueous alteration of igneous materials in geologic units from the Pre-Noachian (>4.1 Ga) and Noachian (4.1-3.7 Ga) periods (Carter *et al.*, 2013; Mustard *et al.*, 2008). The mineral assemblages, chemistry, and geologic setting indicate much of this alteration occurred by water flowing underground (Ehlmann and Edwards, 2014), ranging in depth from shallow sedimentary diagenesis, which depending upon location, comprised acidic, neutral, or alkaline pH fluid (Bristow *et al.*, 2015; Tosca *et al.*, 2005; Yen *et al.*, 2017), to deep, hydrothermal/metamorphic fluid forming serpentine or sub-greenschist facies mineral phases such as prehnite and zeolites (Ehlmann *et al.*, 2009; McSween, 2015). Martian valley networks and open- and closed-basin lake deposits, particularly well-preserved during the Late Noachian and Early Hesperian epochs (3.8-3.3 Ga) (Fassett and Head, 2008; Goudge *et al.*, 2016) also record surface water environments. Rover exploration of sedimentary rocks from two different martian basins revealed shallow playas that experienced multiple episodes of diagenesis by acidic waters (Grotzinger *et al.*, 2005; McLennan *et al.*, 2005) and a Hesperian deep lake with multiple later episodes of groundwater diagenesis and/or hydrothermal alteration, possibly as late as the early Amazonian (~3 Ga) (Grotzinger *et al.*, 2015; Martin *et al.*, 2017; McLennan and al., 2014; Rapin *et al.*, 2018; Yen *et al.*, 2017). Orbital data suggest that other sedimentary basins may have been fed by groundwater (Michalski *et al.*, 2013; Wray *et al.*, 2011), sometimes in communication with magmatic volatiles (Ehlmann *et al.*, 2016; Thollot *et al.*, 2012). Indeed, surface expressions of volcanic hydrothermal or thermal spring systems have been located (Arvidson *et al.*, 2014; Ruff and Farmer, 2016; Skok *et al.*, 2010).

However, after the Late Hesperian (~3 Ga), evidence for liquid water on Mars is sparse. While even young martian meteorites have evidence for aqueous alteration (e.g., (Velbel, 2012), large lava bodies emplaced in the Hesperian and Amazonian do not have hydrated minerals in sufficient abundances to be detectable from orbit (Mustard *et al.*, 2005). Outflow channels, lobate debris aprons, and small valleys occur only near volcanic centers or glacial-like features. Collectively, these data indicate that after a warmer and wetter first 1.0-1.5 billion years, frozen, arid conditions prevailed over the last 3-3.5 billion years (e.g. (Wordsworth, 2016). Alternatively, even the Noachian climate may always have been cold and arid (similar to the last 3-3.5 billion years throughout all of Mars history) with punctuated thermal disturbances to the cryosphere and hydrosphere by the much greater bolide flux (Tornabene *et al.*, 2013).

If martian life emerged, it is possible that it might have looked like the earliest presently-recognized terrestrial record of life, e.g. ~3.4 Ga anoxygenic photosynthesizing microbial mats, forming laminated structures in near-shore, marine facies on a mostly ocean world (Tice, 2009; Tice and Lowe, 2004). However, importantly, martian surface water habitats have always been more episodic and extreme than age-equivalent surface habitats on Earth. All evidence suggests that Earth has had an ocean in continuous existence from at least 3.8 Ga and perhaps from as early as 4.4 Ga (Valley *et al.*, 2002). The preponderance of the martian geological and mineralogical record along with predictions from climate models suggest that no such body of water on Mars was in continuous existence (Carr and Head., 2015; Pan *et al.*, 2017; Wordsworth *et al.*, 2015). Unlike Earth with its stable axial tilt at 23°±1°, Mars' axial tilt fluctuates from 10° to 60° with changes of tens of degrees occurring on timescales of 100's kyr (Laskar *et al.*, 2004). This has driven episodic reorganization of water reservoirs from the poles to mid-latitude belts with concomitant changes in climate cyclically throughout Mars history (Laskar *et al.*, 2002). Occasional flood events from melting of water ice might have caused outflow channels to debouch in the Northern Lowlands of Mars, forming temporary oceans (Tanaka *et al.*, 2003). Certainly, lakes existed for thousands and perhaps millions of years (Fassett and Head, 2008; Grotzinger *et al.*, 2014; Grotzinger *et al.*, 2015). But by the Noachian-Hesperian boundary (~3.7 Ga), the atmosphere was <2 bars thick and possibly only 10s of mbar thick (Bristow *et al.*, 2017; D.Wordsworth *et al.*, 2015; Edwards and Ehlmann, 2015; Hu *et al.*, 2015; Kite *et al.*, 2017; Kite *et al.*, 2014; Wordsworth *et al.*, 2017). Mars had also lost much of its protection from solar radiation and galactic cosmic rays by the loss of its dynamo-driven magnetic field at 3.9-4.1 Ga (Acuña *et al.*, 1999) and the subsequent loss of its atmosphere (e.g. (Ehlmann and al., 2016).

Thus, certainly by ~3.5 Ga and perhaps earlier, Mars' surface environment had evolved to conditions different from and more challenging to life than the time-equivalent habitats on Earth (Westall *et al.*, 2015). Early martian organisms at the surface would have faced at least seasonally sub-freezing temperatures, if not nearly continuous sub-freezing conditions with intermittent thaws, surface aridity, and surface radiation doses many times higher than present on early Earth. The interaction of UV light with Fe and hydrogen peroxide would have produced photo-Fenton chemistry that is lethal to Earth bacteria (Wadsworth and Cockell, 2017). On the other hand, martian subsurface environments with water were widespread and, comparatively, stable. Evidence for groundwater extends to far more recent martian times than that for surface waters and may still be present today (Orosei *et al.*, 2018). An example is the lake in Gale Crater whose sediments are presently being explored by the Curiosity rover. The lake persisted for up to a few million years (Grotzinger *et al.*, 2015), but the sediments bear markers of sedimentary diagenesis long after the lake had vanished. Cross-cutting geologic relationships show that at least several tens of meters of lake sediment had to be eroded, overlain by dunes, the dunes lithified to sandstone, and then crosscut by diagenetic sulfate and silica veins in multiple generations of subsurface fluid flow, persisting even into the Amazonian (Frydenvang *et al.*, 2017; Rampe *et al.*, 2017; Yen *et al.*, 2017). Fracture networks provided a conduit between habitable subsurface aquifers and more transient surficial habitable systems. Elsewhere, fluid circulation through deep fracture networks driven by hydrothermal activity within impact craters also mobilized fluids from the surface to far beneath the cryosphere (Osinski *et al.*, 2013).

Lastly, an important difference between habitable environments on Earth and Mars may be related to differences in communication between the surface and subsurface. Whereas on Earth, warm temperatures and abundant liquid water provided a rapid pathway for recolonizing the surface from the subsurface after extinction events (e.g. impact events) (Abramov and Mojzsis, 2009) or global glaciation, on Mars sub-freezing surface temperatures and the formation of a thick, global permafrost layer (i.e. cryosphere) might have limited communication between surface and subsurface habitats, particularly later in Mars history (Clifford, 1993; Clifford *et al.*, 2010; Clifford and Parker, 2001; Grimm *et al.*, 2017; Harrison and Grimm, 2009). Therefore if periodic warm conditions did occur at the surface, the pathways for communication with subsurface may not have been as easily established for (re)colonization of the martian surface during brief Hesperian surface habitable periods.

Consequently, rock-hosted habitats showing evidence of persistent water warrant considerable attention in the search for martian life (Westall *et al.*, 2015). Some of these systems may have been uninhabitable, perhaps challenged by salinity and acidity (Tosca *et al.*, 2008). Nonetheless, the most globally widespread systems and some sites explored in situ are marked by neutral to alkaline waters of low salinity, which, if on Earth today would be habitable (Ehlmann *et al.*, 2011; Grotzinger *et al.*, 2015).

Several candidate martian landing sites under consideration for future exploration missions have accessible stratigraphy that may preserve rock-hosted habitats. These include aquifers in volcanic rock and in sedimentary rock. Most immediately, the volcanic rock aquifer with clay minerals, carbonate, and serpentine exposed by erosion at Northeast Syrtis is under consideration for the Mars-2020 rover mission. These ancient habitats can and should be explored at high priority as available habitats for martian life, using the lessons and strategies derived from the terrestrial modern and paleorecords of the quantities, nature, and locations of biosignatures of past rock-hosted life.

## 3. Modern Rock-Hosted Life On Earth
### 3.1. Geologic Settings with Rock-Hosted Life

Microbial communities have been detected globally in a wide variety of rock-hosted environments. Their abundance and community structure reflect physicochemical properties of the rock/water host, the type and rates of energy and nutrient fluxes, geo-biological feedback, and the geological history of the rock. Though most rock-hosted life are comprised of Archaea and Bacteria, active eukaryotic members exist. These range from protists in deep aquifers (Sinclair and Ghiorse, 1989) to fungi in sub-seafloor sediment (Orsi *et al.*, 2013; Pachiadaki *et al.*) and 793-meter deep fracture waters in granite (Sohlberg *et al.*, 2015) to multicellular bacteriophagous nematodes (Borgonie *et al.*, 2011). What follows is a very brief survey of the variety of rock-hosted ecosystems documented on Earth, some of which could represent terrestrial analogs to potential martian rock-hosted ecosystems (Fig. 1).

The shallowest examples of a rock-hosted ecosystem are the highly concentrated cryptoendolithic communities existing millimeters beneath the rock surface, which are not truly "rock-hosted life" as we define the term here (see Introduction). The primary producers of these communities, cyanobacteria and algae, are surface dwelling photosynthesizing organisms that have retreated to the near subsurface to reduce their exposure to moisture and temperature extremes while retaining access to a sustainable photon flux (Friedmann, 1982; Wong *et al.*, 2010). Some shallow subsurface ecosystems do use their rock/soil hosts for metabolism, meeting our definition of rock-hosted life. Chemoautotrophic aerobic and anaerobic microorganisms that fix atmospheric $CO_2$ also reside in barren polar soils and metabolize atmospheric trace gases such as $H_2$, CO (Ji *et al.*, 2017) and $CH_4$ (Edwards *et al.*, 2017; Lau *et al.*, 2015). Though there is a significant energetic potential for such metabolisms in martian regolith, especially utilizing CO, there is no detectable presence of this metabolism yet (Weiss *et al.*, 2000; Yung *et al.*, 2018).

For the majority of the continental surface on Earth, heterotrophic bacteria involved in the degradation of organic photosynthate (e.g. cellulose) dominate soil communities (Federle *et al.*, 1986) and shallow aquifers (Balkwill and Ghiorse, 1985). Similarly, organic detritus input from the sea surface, water-column or continental photosynthates dominate the reductant input of shallow sub-seafloor sediments (Fig. 1). Sediment pore waters host a wide variety of specialized microbes that generally use just one of a succession of oxidants present in the system to degrade organic matter, starting with aerobic oxidation at the sediment/water interface or at the aquifer recharge zone. This succession is controlled by which oxidant offers the greatest release of free energy from the process; in order they are dissolved $O_2$ >Mn oxides >nitrate >Fe oxides >sulfate (Froelich *et al.*, 1979) (Fig. 2). The reason for this ordered succession is that the group of microorganisms can utilize the energy released for growth and can thus outcompete those that receive less energy. Kinetic factors can also shape community metabolism and influence which molecular forms of given electron donors or acceptors are used (Bonneville *et al.*, 2004). The oxidized by microbial metabolic processes may have a distinctive isotopic signature and may form carbonate minerals with compositions that reflect the oxidant used (e.g., reduced manganese carbonate, rhodochrosite) (Coleman *et al.*, 1982). The carbon isotope composition of photosynthetic organic matter is distinctive with both terrestrial and marine derived material having a much more negative $\delta^{13}C$ value than dissolved marine carbonate or the open-water calcium carbonates precipitated from it. The materials precipitated in sediment pore space by the organic matter degradation (mineralization) processes retain this distinctive isotopic characterization to a greater or lesser extent. That extent is governed by the amount of admixture with ocean water-dissolved carbonate, which plays a more significant role nearer to the sediment-water interface where there is faster diffusive access to overlying water column. In organic-rich sediments these processes continue in the pore waters with burial until the organic oxidants are exhausted.

In the lowermost oxidation zone, where the least exergonic electron acceptor remains, methanogenic Archaea operate on the residual organic matter, or $CO_2$ and subsurface $H_2$ (Fig. 2). Various metabolic processes all produce $^{13}C$-depleted $CH_4$ and leave residual bicarbonate equivalently isotopically enriched. Again, this bicarbonate may be precipitated as distinctive carbonate minerals sometimes with contributions from the bacterial processes mentioned above. The $CH_4$ may diffuse upward and itself be oxidized anaerobically by archaea, bacteria and archaeal-bacterial consortia using sulfate or other oxidants in shallow sub-seafloor sediment (Cai *et al.*, 2018; Ettwig *et al.*, 2010; Haroon *et al.*, 2013; Kits *et al.*, 2015; Milucka *et al.*, 2012; Orphan *et al.*, 2001). At greater depths and temperatures exceeding 100°C, any residual organic photosynthate is thermally matured and may be transformed to oil and/or gas or coal while the host strata become sterilized of indigenous microorganisms, a process described as "paleopasteurization" (Wilhelms *et al.*, 2001). The microbial communities associated with oil, gas or coal endowed deposits either represent immigrants arriving with groundwater flow over geological time as the deposits cooled below their maximum temperatures (Tseng *et al.*, 1998), represent residents indigenous to the sandstone reservoir when oil or gas migrated upwards to be trapped (Wilhelms *et al.*, 2001), or represent contaminants introduced during flooding of or production from the oil reservoirs (Dahle *et al.*, 2008). Although these sediment-hosted microorganisms are living completely or nearly completely off the detritus of photosynthetate, it is likely that any organic matter from abiotic chemosynthetic or biotic chemolithoautotrophic sources could be processed in a similar fashion. The significance of these organic degradation processes is that the inorganic metabolic products may produce characteristic mineral phases, and isotopic and chemical signatures. These can be durable biosignatures that survive for very long periods of time (see Section 4, *Biosignatures Of Past Rock-Hosted Life*).

In continental environments where the water table is hundreds of meters deep (Fig. 1), subsurface communities rely upon chemolithotrophs living off atmospheric or vadose zone gases (Jones *et al.*, 2016; Tebo *et al.*, 2015; Webster *et al.*, 2016), redox reactions with reduced minerals (Mansor *et al.*, 2018), and metals present in carbonate (Barton and Northup, 2007). The chemolithotrophs serve as primary producers for complex communities (Dattagupta *et al.*, 2009; Fraser *et al.*, 2017) and as ecosystem engineers excavating large caverns by dissolution of carbonate and deposition of sulfate by the sulfuric acid they produce (Mansor *et al.*, 2018). In water-saturated environments where oxygenated water penetrates deeply into crustal rock such as mountainous terrains of North America (Murdoch *et al.*, 2012; Osburn *et al.*, 2014; Sahl *et al.*, 2008) or the basaltic flows of the geothermal environments of Iceland (Trias *et al.*, 2017) or through taliks in 500 meter thick permafrost into underlying Archean metamorphic rock (Onstott *et al.*, 2009), heterogeneous redox conditions create highly exergonic conditions for S, Fe, N, and Mn oxidation and subsurface microbial communities are dominated by chemolithotrophic primary producers. Where oxygenated seawater comes into contact with marine basaltic crust, chemolithoautotrophs are the primary producers fixing $CO_2$ to support substantial biomass by mediating electron transfer at mineralogical redox interfaces from reduced forms of Fe, S, and Mn to the aerobic fluids in pore spaces (Edwards *et al.*, 2012). Unlike the organic-rich sediments discussed above, the geochemical evidence suggests that these rock-hosted communities do not rely upon the groundwater transport of organic photosynthate (Kieft *et al.*, 2018) (Fig. 2).

Of the bioavailable electron donors being utilized by deep, water-saturated, rock-hosted communities, $H_2$ is probably a key fuel in the deep biosphere (Nealson *et al.*, 2005) and several abiotic modes of formation exist within rock-hosted environments. In anaerobic volcanic aquifers (Fig. 1), basalt interacts with anaerobic groundwater releasing $H_2$, which then supports chemolithotrophic microbial communities are supported by the oxidation of $H_2$ at depths of hundreds of meters (Mayhew *et al.*, 2013; Stevens and McKinley, 1995). Even in the absence of $O_2$, anaerobic Fe-oxidation via nitrate reduction is a metabolic process that can recycle $Fe^{2+}$ produced by $Fe^{3+}$ reduction that can lead to a subsurface Fe-cycle, depending upon the availability of nitrate (Melton *et al.*, 2014). Given the presence of nitrogen-bearing compounds – including nitrate – (Stern *et al.*, 2015) and the Fe-rich nature of the martian crust, the presence of such a metabolic network has been proposed for the martian subsurface (Price *et al.*, 2018).

Radiolysis of groundwater also generates $H_2$, $H_2O_2$ and $O_2$, that has been shown to sustain subsurface chemolithoautrophic primary producers by providing not only $H_2$ as an electron donor, but also electron acceptors, such as sulfate via oxidation of sulfides by radiolytically produced $H_2O_2$ (Lefticariu *et al.*, 2006; Li *et al.*, 2016; Lin *et al.*, 2006). Metagenomic analyses combined with metaproteomic and metatranscriptomic analyses have also revealed that within these radiolytically supported communities, a dynamic and temporally varying multi-tier energy pyramid of chemolithoautotrophs exists that recycles biogenic $CH_4$ and sulfide and possibly nitrogen, while fixing $CO_2$ using the Wood-Ljungdahl pathway and Calvin-Benson-Bassham cycle (Lau *et al.*, 2016; Magnabosco *et al.*, 2015; Magnabosco *et al.*, 2018c). The bacterial biomass supports multicellular bacteriophagous nematodes at the top of the food chain (Borgonie *et al.*, 2011). These results indicate that radiolysis combined with commensurate syntrophic interactions constantly recharge the redox couplings in these environments and that they are not chemically stagnant as claimed by McMahon *et al*. (McMahon *et al.*, 2018). The radiolytic $H_2$ production rate on Mars is just as great as that found in the crustal rocks of Earth despite the lower concentrations of radiogenic isotopes, primarily because of the likely higher porosity at a given depth due to the lower gravity on Mars (Dzaugis *et al.*, 2018; Onstott *et al.*, 2006; Tarnas *et al.*, 2018).

Cataclastic diminution of silicate minerals in the presence of water can also generate $H_2$ (Kita *et al.*, 1982) and in the presence of $CO_2$, generate CO and $O_3$ (Baragiola *et al.*, 2011). $H_2$ release during seismic events has been recorded at 3 km depths in South Africa (Lippmann-Pipke *et al.*, 2011) and $H_2$ release during rock-crushing at the base on the 3 km thick Greenland ice sheet has been inferred (Telling *et al.*, 2015). The relationship between rock fracturing and/or crushing and subsurface microbial community abundance and activity is not yet resolved and is an avenue of current research in subsurface microbiology. Nonetheless, its implications for subsurface life on Mars, which has fracturing due to impacts and tectonics, have already been proposed (McMahon *et al.*, 2016).

At still greater depths and at temperatures >200ºC in peridotite, serpentinization produces abundant $H_2$ in high pH fluids (McCollom and Bach, 2009). This $H_2$ as well as that generated by radiolysis in turn reacts with transition metal sulfide catalysts to produce $CH_4$ and low molecular weight hydrocarbons via Fischer-Tropsch-type synthesis (McCollom, 2016; Sherwood Lollar *et al.*, 2006; Sherwood Lollar *et al.*, 2002). The resulting hydrocarbons can either diffuse upwards to support chemolithotrophs, methanotrophs, and heterotrophic, alkane-degrading anaerobic bacteria at shallower, cooler temperatures or remain trapped until the host-rock has cooled down whereupon these microbial metabolic clades can penetrate the serpentinite during groundwater flow and utilize the hydrocarbons (Purkamo *et al.*, 2015). These processes that support the subsurface microbial communities found in continental ophiolite complexes, such as the Samail Ophiolite (Rempfert *et al.*, 2017), in metamorphosed komatiites of Archean greenstone belts (Sherwood Lollar *et al.*, 2005), have also been hypothesized to support a martian subsurface biosphere (Schulte *et al.*, 2006; Westall *et al.*, 2013).

These represent a few examples of the types of rock-hosted microbial communities that are globally distributed across the Earth at depths ranging from millimeters to kilometers in a wide range of rock types and that are metabolically and phylogenetically diverse (Mykytczuk *et al.*, 2013). In general, more oxic conditions nearer to the surface yield to more reduced conditions with increasing depth but with important exceptions. Despite the great abundance of organic carbon derived from the surface photosphere in marine sediments and shallow soils, chemolithotrophy is widespread and even dominant in many subsurface environments. This may explain the absence of any correlation of deep subsurface prokaryotic biomass with organic carbon content in the continental subsurface below the soil zone (Magnabosco *et al.*, 2018a).

*3.2. Fundamental Physical and Environmental Controls on Rock-Hosted Life*

The thermal state of the crust constrains the habitable zone. The currently recognized temperature limits for metabolic activity range from -20°C for microorganisms trapped in Siberian permafrost (Rivkina *et al.*, 2000) and -25°C for an aerobic, halophilic heterotroph in laboratory microcosm experiments utilizing $^{14}C$-labeled acetate (Mykytczuk *et al.*, 2013) up to 122°C for a methanogen isolated from a deep sea vent plume (Takai *et al.*, 2008). Based upon temperature alone, the habitable volume for the Earth's continental and oceanic crust has been estimated to be ~$2\times10^{18}$ $m^3$ (Heberling *et al.*, 2010; Magnabosco *et al.*, 2018a), using global heat flow and surface temperature maps and thermal conductivity estimates. In the case of the continental crust the average depth to the 122°C isotherm is 4 km; a maximum depth of 16 to 23 km occurs in the Siberian Craton where mean annual temperatures and heat flow are both lower than average (Magnabosco *et al.*, 2018a). Similar types of calculations for Noachian Mars indicate an average depth to the 122°C isotherm would have been 6-8 km and the corresponding habitable volume based on temperature constraints alone would also be ~$10^{18}$ m$^3$ (Michalski *et al.*, 2017).

Temperature and possibly the ionic strength of the crustal fluids play a role in constraining the abundance and activity of subsurface life since cellular concentrations appear to be inversely correlated with both parameters (Magnabosco *et al.*, 2018a). Organic markers of biodegradation of petroleum suggest that the maximum temperature of the subsurface biosphere may typically be closer to 80-85°C and that salinity >50 g L$^{-1}$ may inhibit low energy metabolisms such as methanogenesis to even lower temperatures (Head *et al.*, 2014). High concentrations (>220 g L$^{-1}$) of chaotropic salts, such as $MgCl_2$, may even preclude life (Hallsworth *et al.*, 2007).

The fluid-bearing (saturated or thin film) porosity that is accessible, i.e. with pore throats that are greater than 0.1 μm in diameter, also controls the habitable volume. On Earth rock strata from 3 to 5 km depth may have a matrix porosity of 0.5 to 1%, but their habitable volume could be as little as 0.05 to 0.002% (Supplementary Fig. S1) due to compaction and cementation. On Mars the porosity is likely to be 10% at comparable depths due to the lower gravitational force and as a result the subsurface habitable volume on Mars may be greater than that of the Earth's.

Porosity and permeability also constrain the flux of nutrients and the degree of metabolic activity. For terrestrial life, a finite minimum quantum of energy determined by the reaction ADP + P => ATP must be available through catabolic redox or substrate level reactions to be metabolically useful (Müller and Hess, 2017). To sustain life, the Gibbs free energy flux (energy per unit time per cell) (Hoehler, 2004; Onstott, 2004) must be equal to or exceed that required for a cell's (Hoehler and Jørgensen, 2013; Onstott *et al.*, 2014) or a syntrophic community's (Scholten and Conrad, 2000) maintenance. Temperature is a principal control on the maintenance energy demand in part because of the diffusivity of H$^+$ through the cell membranes increases with temperature (van de Vossenberg *et al.*, 1995). As a result cell metabolic rates must increase with temperature to counteract these effects. The higher the temperature, the higher the nutrient flux needed to maintain a given subsurface biomass. The same may also hold true for salinity as microorganisms need to manufacture internal osmolytes to maintain osmotic pressure and osmolytic production exerts an additional energy requirement (Oren, 1999).

To the extent that higher rock permeability increases groundwater velocities that increase the rate at which reactants flow towards microorganisms hosted by rock strata, higher permeability can lead to a more prolific subsurface biomass by maintaining a non-zero Gibbs free energy (Marlow *et al.*, 2014a). Three other abiotic processes that operate to enhance the local flux of nutrients are chemical gradients or boundaries, physical heterogeneity, and local abiotic and biotic recycling. First, the presence of high electron donor/acceptor chemical spatial gradients in rock units enhance local diffusive fluxes which lead to higher microbial activity and biomass than in homogenous units. On Earth these are typically found at the contacts between organic-rich shale and sulfate-bearing sandstone (Krumholz *et al.*, 1997) or oil and water (Bennett *et al.*, 2013) producing a higher rates of microbial metabolism than observed some distance away from these contacts. On Mars an example might be the boundary between an olivine-rich clay-bearing unit and an overlying sulfate-rich unit in NE Syrtis (Marlow *et al.*, 2014a). Serpentinization of the olivine would generate $H_2$ that would then diffuse into the overlying aquifer where it could be microbially reduced by sulfate, leading to a high biomass at the boundary. Second, physical heterogeneity can also act to create favorable zones, as is observed in the high cell concentrations within highly fractured and brecciated rock of the Chesapeake Bay impact structure compared to the overlying marine sediments (Cockell *et al.*, 2012). In this case metabolically active microorganisms are constrained to the fractured rock where they draw down the energy substrates and increase the product concentrations. Though the surrounding massive rock has pore spaces too small for microbes, the diffusive flux of energy substrates from the massive rock into the fractured zone and the diffusion of the products from the fractured zone into the massive rock enhances the biomass residing in the fracture rock (Supplementary Fig. 1). Third, abiotic recycling reactions such as radiolysis can continuously generate $H_2$ from water and recycle metabolic waste products, such as $HS^-$ back into sulfate and sustain subsurface microbial communities without the need for fluid transport (Lin *et al.*, 2006). "Cryptic sulfur cycling", in which iron oxides abiotically oxidize sulfide to more oxidized sulfur species, can support organic carbon degradation in non-stoichiometric proportions with a relatively limited sulfur supply (Holmkvist *et al.*, 2011). Finally, syntrophic interactions between different microorganisms also act to sustain subsurface communities by converting waste products back into reactants locally without the need for advective transport (Lau *et al.*, 2016). Thus, obligately mutualistic metabolism (Morris *et al.*, 2013) may be characteristic aspect of subsurface microbial communities as a means of avoiding extinctions (Gaidos *et al.*, 1999) because Gibbs Free Energy will remain nonzero.

*3.3. Biomass Distribution*

Magnabosco *et al.* (2018a) recently estimated the total living subsurface prokaryote biomass for the Earth was $7-11 \times 10^{29}$ cells of which $2-6 \times 10^{29}$ cells occur in the continental subsurface, $2 \times 10^{29}$ cells exist the oceanic crust and $3 \times 10^{29}$ cells reside in the sub-seafloor sediments. Permafrost-affected crust and continental ice sheets cover a large fraction of the continental area and are often considered terrestrial analogs to early Mars. The total cellular concentrations for Arctic permafrost sediments can be quite high ranging from $10^7$ to $10^9$ cells $gm^{-1}$ (Fig. 3A) and diminish with depth and increasing permafrost age up to 2 Ma (Gilichinsky and Rivkina, 2011). The cell concentrations within the Greenland ice sheet are much lower, on the order of $10^5$ cells $cm^{-3}$, except at the very bottom where the ice sheet is in contact with the ground where cell concentrations reach $10^9$ cells $cm^{-3}$ (Fig. 3A). Permafrost and subglacial sediments from Antarctica exhibit cell concentrations that are also greater than those of the ice sheet. In general, cell concentrations in ice sheets do not diminish as a function of depth and age and likely reflect a combination of airborne input flux and in situ metabolism (Chen *et al.*, 2016).

The cell concentrations for suspended cells in groundwater exhibit little decline with depth, overlap the concentrations observed in ice cores, averaging around $10^5$ cells cm$^{-3}$, and exhibit a broad range from $10^1$ to $10^9$ cells cm$^{-3}$ (Fig. 3B). The cell concentrations from rock and sediment cores unlike ice and groundwater samples decline with increasing depth following a power law fit (Fig. 3C). Notable exceptions are the cellular concentrations reported for the Chesapeake Bay Impact that increase at a depth of 1.5 km where the highly fractured basement rock exists (Cockell *et al.*, 2012). In soil zones the concentrations range from $10^9$ cells gm$^{-1}$ down to as low as $10^{5-6}$ cells gm$^{-1}$ in the case of the Atacama desert (Connon *et al.*, 2007; Lester *et al.*, 2007), considered by some to be a terrestrial analog site for Mars because of its aridity and low organic content. Below 10 meters depth, however, cell concentrations do not correlate with water saturation of the pore space. For example, the cell concentrations within unsaturated volcanic ash deposits at a depth of 400 m in central Nevada (deep vadose zone) range from $5 \times 10^4$ to $5 \times 10^7$ cells g$^{-1}$ (Haldeman and Amy, 1993a; Haldeman *et al.*, 1993b), which is not significantly different from the cell concentrations reported for water saturated post-impact sediments of the Chesapeake Bay Impact (Breuker *et al.*, 2011; Cockell *et al.*, 2012) or Atlantic Coastal Plain Sediments (Magnabosco *et al.*, 2018a) of the same depth (Fig. 3C).

The cell concentrations also do not correlate with the rock type, i.e., sedimentary versus igneous versus metamorphic, with the possible exception of salt deposits where cell concentrations range from 0.02 to $10^4$ cells g$^{-1}$ (Schubert *et al.*, 2009b; Schubert *et al.*, 2009a; Schubert *et al.*, 2010; Wang *et al.*, 2016). Despite their paucity within salt microorganisms exhibit remarkable preservation with viable cells being isolated from salt deposited tens of thousands to hundreds of millions of year in the past (Jaakkola *et al.*, 2016), though the older claims remain controversial (Hebsgaard *et al.*, 2005; Lowenstein *et al.*, 2005).

Cell concentrations on fracture or cavity surfaces are often considerably higher than those of the surrounding fluid or matrix especially if the interface acts to focus redox fluxes. For example, in deep vadose zones cell concentrations up to $10^7$ to $10^8$ cell/cm$^2$ and prolific and pigmented biofilms exist on the surfaces of caverns in quartzite (Barton *et al.*, 2014), basalt (Riquelme *et al.*, 2015) and carbonate (Jones *et al.*, 2016). Because of the difficulty of aseptically sampling water saturated fracture surfaces at depth only two studies of the cell concentrations on deep fractures have been published. Analysis of modern biofilms, occurring on fracture surfaces in 2.7 Ga metavolcanic rocks at a depth of 2.8 km, revealed $10^5$ cells/cm$^2$ with cells occurring in clumps of 2 to >20 (Wanger *et al.*, 2006). Given the fracture width, such a concentration corresponded to a 100x enhancement of the living cell concentration relative to the fracture water. Much lower living cell concentrations, 40 to $2 \times 10^3$ cells cm$^{-2}$, have been reported for 186 m deep groundwater-fed fractures in granite (Jägevall *et al.*, 2011). Although these two studies would suggest that deep fracture surfaces do not harbor high biomass concentrations, examination of buried Cretaceous hydrothermal veins reveals preserved organic remains of microbial colonies in mineral surfaces which by mass would be equivalent to ~$10^9$ cells $cm^{-2}$ (Klein *et al.*, 2015).

McMahon et al. (2018) stated "the bulk of Earth's massive deep biosphere, and presumably also its fossil record, is a poor analog for any ancient or modern Martian equivalent which, in the absence of a productive surface biosphere, would be much smaller and dominated by chemoautotrophs, not heterotrophs". This statement appears is not borne out by existing data. As described above, the cell concentrations in continental rock and groundwater do not exhibit any correlation with dissolved or particulate organic carbon concentrations. This finding contrasts with the observations of shallow sub-seafloor sediments where cell concentrations do correlate with the organic photosynthate content (Lipp *et al.*, 2008). Yet the overlap in cellular concentrations of the continental rocks with those of deep sub-seafloor sediments suggest that access to organic photosynthate has little impact on deep subsurface biomass (Fig. 3C). The deep subsurface environments associated with Phanerozoic age oil (Head *et al.*, 2014) and coal (Kirk *et al.*, 2015) deposits where heterotrophic metabolisms would perhaps dominate comprise only $10^{12}$ $m^3$, or 0.0001%, of the total habitable volume of the Earth's subsurface biosphere. The cell abundance data and observed metabolisms do not support the claim that most of Earth's deep biosphere is sustained by heterotrophic metabolism of surface derived photosynthate. Rather, fracture surface concentrations of $10^5$ to $10^9$ cells/$cm^2$ are observed in Earth's chemolithoautotrophic communities in settings isolated from organic photosynthate and fueled by chemolithoautrophy, which is an appropriate analog to Mars. (See section 4 for a discussion of Earth's fossil record.)

Using simple assumptions from the chemical energy available from basalt weathering ($10^{-13}$ kJ/gm-yr), Jakosky and Shock (Jakosky and Shock, 1998) estimated that over 4 billion years Mars could accumulate 10 grams $cm^{-2}$ of biomass from a 100 meter thick basalt layer. For comparison, Earth's continental crust is estimated to contain 0.006 to 0.02 grams of extant life $cm^{-2}$ integrated from 1 meter depth to the 122$^o$C isotherm (Magnabosco *et al.*, 2018a). The addition of fluid flow from an oxic surface to reducing subsurface, as where sulfate rocks are in contact with serpentinizing rocks, increases this estimate by an additional 10 grams of biomass $cm^{-2}$ by enhancing the delivery of reactants and removal products (Marlow *et al.*, 2014a). Calculations of the energy flux from subsurface radiolytic reactions for Mars reveal an energy source comparable to that found on Earth indicating that the subsurface biomass abundance should be comparable to that of the Earth's (Dzaugis *et al.*, 2018; Onstott *et al.*, 2006) and allay the concerns raised about limited oxidant supply to the martian subsurface from the surface (Fisk and Giovannoni, 1999). Radiolysis releases energy into the rock at a rate of $10^{-9}$ kJ/gm-yr based upon the parameters utilized by (Onstott *et al.*, 2006) of which some fraction is accessible for biomass production depending upon the porosity. This rate is greater than that estimated for weathering reactions and would be even higher on Mars during the Noachian when the radioactive parent isotopes were more abundant. These calculations suggest that adequate energy exists to support substantial biomass and accumulate organic matter over time in subsurface aquifers, independent of a habitable surface world.

*3.4. Biodiversity and Geographic Scaling*

As is the case for biomass, the species richness of deep subsurface environments is highly variable. Reports of deep subsurface planktonic communities dominated by a single archaeal (Chapelle *et al.*, 2002) or bacterial (Chivian *et al.*, 2008) species are rare. More commonly diversity estimates range from over a hundred (Marteinsson *et al.*, 2013) to almost 100,000 (Bomberg *et al.*, 2016) operational taxonomic units or OTUs (at 97% identity in the 16S rRNA gene) within a single fluid sample. To some extent this reflects the improvement in sequencing technology and the fact that the highly variable sub-regions of the 16S rRNA gene are targeted. It is not unusual to find a large number of OTUs that comprise <1% of the total population (Castelle *et al.*, 2013; Magnabosco *et al.*, 2014), referred to as the rare biosphere (Sogin *et al.*, 2006). Additionally, single cell genome sequencing of subsurface *Ca.* Desulforudis audaxviator indicates significant differences in the genomes of single species (Labonté *et al.*, 2015).

Species abundance does not correlate with its ability to influence the overall function of a subsurface community. For example, in continental subsurface environments, methanogens frequently comprise only two percent of the total community, but the primary gas phase is biogenic $CH_4$. In the case of one deep subsurface site in South Africa this biogenic $CH_4$ was the principal carbon source for the remainder of the community (Lau *et al.*, 2016; Simkus *et al.*, 2015). Electron microscopy (Engelhardt *et al.*, 2014; Kyle *et al.*, 2008; Middelboe *et al.*, 2011), single cell genomes (Labonté *et al.*, 2015), and metatranscriptomic analyses (Lau *et al.*, 2016) have also revealed that viruses are abundant and actively infecting bacteria and altering their genomes (Paul *et al.*, 2015) in sub-seafloor sediments and fractured rock aquifers. Because active viral populations can transfer genes between microbial species and control the population density they may be a vital component of obligately mutualistic metabolic SLiMEs.

One might expect that subsurface diversity would resemble island-like behavior because of the lesser connectedness between subsurface habitats when compared to surface habitats where wind and surface water are transport agents. In island-like ecosystems the number of species should increase with the size of the island (i.e. volume of groundwater sampled) (Locey and Lennon, 2015). However, Magnabosco *et al.* (2018a) did not find any such correlation. Species richness may instead correlate with greater heterogeneity of microenvironments, which are difficult to characterize in the subsurface. The lack of species richness versus habitat size could also reflect a surprisingly high degree of connectedness between habitats. This is consistent with the presence of the same rare biosphere OTU's in fractures ranging from 0.6 to 3.0 km depth and separated by hundreds of kilometers in South Africa and a general lack of a distance-decay relationship (Magnabosco *et al.*, 2018a). The presence of the same OTU's in fracture water of different isotopic compositions also suggests that these species are actively motile (Magnabosco *et al.*, 2014). The implication of these observations for an early martian subsurface biosphere it that impacts and volcanic activity could have produced zones of sterilized rock, but these zones would have quickly become recolonized by groundwater circulation.

*3.5. Subsurface Metabolic Activity*

Estimates of the *in situ* metabolic rates of subsurface ecosystems offer insight into their longevity and biomass turnover and provide a better sense of how they impact biogeochemical cycles on a planetary scale, though obtaining accurate estimates has proven challenging (Orcutt *et al.*, 2013). Geochemical estimates of the electron production rate from marine subsurface *in situ* microbial activity range from ~5 mols $e^- L^{-1} yr^{-1}$ at seafloor hydrothermal vents (Wankel *et al.*, 2011) to ~2 to 100 pmols $e^- L^{-1} yr^{-1}$ in oligotrophic sub-seafloor red clays (Røy *et al.*, 2012) to ~4 to 0.004 pmols $e^- L^{-1} yr^{-1}$ for continental deep fractured rocks and consolidated sediments (Kieft and Phelps, 1997), estimates spanning more than 15 orders of magnitude. Recent metabolism-agnostic approaches utilize isotopic labeling and measurements of the D/L of aspartic acid of bulk sub-seafloor sediment (Lomstein *et al.*, 2012) and of cells separated from deep continental fracture fluids (Onstott *et al.*, 2014) to determine the bulk rate of growth and repair often referred to as cell turnover . These analyses yielded cell turnover times ranging from 73,000 years for sub-seafloor sediments to several months for 60ºC fracture water and imply that *in situ* metabolic rates are strongly temperature dependent (Onstott *et al.*, 2014; Trembath-Reichert *et al.*, 2017; Xie *et al.*, 2012). The shorter cell turnover times observed in deep continental fracture water are also consistent with metatranscriptome and metaproteome observations from similar environments that revealed significant intra-community recycling of metabolic waste products (Lau *et al.*, 2016). Recycling of biogenic $CH_4$, sulfide and $CO_2$ suggest that initial geochemical approaches for estimating metabolic rates (Phelps *et al.*, 1994) may have underestimated the rates of cell turnover. Despite the challenges posed in accurately estimating the metabolic rates of subsurface microbial ecosystems, recent *in situ* approaches utilizing natural $^{14}C$ (Simkus *et al.*, 2015) suggest that a wide range can be found and correlates with the energy fluxes and temperatures. The longer turnover times in colder ecosystems raise questions about the nature and rate of evolution in subsurface ecosystems, which are also characterized by physicochemical conditions that are more stable over time than surface ecosystems and where microorganisms are exposed to low radiation dosage rates relative to those of surface ecosystems (Teodoro *et al.*, 2018).

*3.6. Evolution of Chemoautotrophic versus Photosynthetic Pathways*

The depiction of subsurface microbial environments above as dominantly suboxic conditions near the surface yielding to more reduced conditions with increasing depth is the result of the evolutionary emergence of oxygenic Photosystem II in cyanobacteria. Prior to this emergence both surface and subsurface habitats were likely dominated by anaerobic metabolisms such as methanogenesis. Recently retrieved genomes of methanogens belonging to Crenarchaeota have revised our understanding of when methanogenic metabolisms evolved. The discovery of Verstraetearchaeota (Vanwonterghem *et al.*, 2016) and Bathyarchaeota (Evans *et al.*, 2015) indicate that the methanogenic metabolic pathway must have arisen after the split between Archaea and Bacteria from the Last Universal Common Ancestor, LUCA, but prior to the split between the Crenarchaeota and Euryarchaeota. This places the time for the emergence of this pathway in the Paleoarchean or Hadean. The youngest bound on age for the development of the methanogenic pathway is ~3.25 Ga, i.e., prior to, or during the proposed Archean expansion of genes, based upon molecular clock analyses (David and Alm, 2011) and likely developed in the stem of Archaea (Betts *et al.*, 2018).

Recently sequenced genomes from non-photosynthetic members of the Cyanobacteria phylum indicate that oxygenic photosynthesis arose within Cyanobacteria after the split of photosynthetic Cyanobacteria (now the class of Oxyphotobacteria) from the other non-photosynthetic Cyanobacterial lineages, Melainabacteria and Sericytochromatia (Soo *et al.*, 2017). The age for this divergence has been estimated to be 2.5 to 2.6 Ga based upon molecular clocks (Shih *et al.*, 2017), consistent with the ~2.8 Ga age for the stem of the Cyanobacteria to 2.2 Ga time interval required for the evolution of oxygenic photosynthesis estimated from a different molecular clock approach (Magnabosco *et al.*, 2018b). This is consistent with the theory that Photosystem II evolved from a Mn-carbonate oxidizing enzyme within a suboxic, neutral pH paleoocean (Johnson *et al.*, 2013) to a $HCO_3^-$-oxidizing oxygenic photosystem and then to $H_2O$-oxygenic photosynthesis (Dismukes *et al.*, 2000) just prior to the rise of $O_2$ during the Great Oxidation Event, GOE, at ca. 2.3 Ga (Betts *et al.*, 2018). During this transition, surface anaerobic ecosystems would have either started going extinct or would have adapted to higher $O_2$ levels perhaps incorporating aerobic metabolic pathways whereas deep subsurface ecosystems would have remained relatively unaffected.

The molecular clock constraints on the non-oxygenic phototrophic bearing phyla, Chloroflexi (green non-sulfur bacteria) and Chlorobi (green sulfur bacteria) suggest that their stems date from no earlier than ~3 Ga with $Fe^{2+}$-oxidizing GSB likely being the most ancient (Magnabosco *et al.*, 2018b). The inferred $Fe^{2+}$ phototrophic mat structures in the 3.45 Ga Buck Reef Chert (Tice and Lowe, 2004) predate this stem age for all phototrophic lineages. The remaining phototrophic bacteria within the phyla Proteobacteria and Clostridia likely acquired their abilities by later horizontal gene transfer.

A critical nutrient to the expansion of both subsurface and surface life on any planet is the availability of nitrogen as an aqueous species. On Earth, microorganisms evolved the ability to fix $N_2$ into ammonia with the development of nitrogenase to overcome this constraint. Nitrogenases, Nif proteins, are complex enzymes, utilizing iron, molybdenum and/or vanadium that exist in both bacterial and archaeal domains. Phylogenetic comparison of genes that comprise nitrogenases and a complement of proteins required for their regulation indicate that nitrogenases emerged in anoxic sulfidic environments on the Earth within obligate anaerobic thermophilic methanogens and were transferred to obligate anaerobic clostridia (Boyd *et al.*, 2015), both common subsurface microorganisms. As Nif proteins were adopted first by the aerobic diazotrophic lineage Actinobacteria and then by the more recently evolved aerobic Proteobacteria and Cyanobacteria lineages, the Nif protein suite became more complex to protect the core MoFe-bearing proteins from $O_2$ (Boyd *et al.*, 2015). Although it is not clear whether the emergence of the more complex protein occurred prior to or after the GOE, it is certain that the ancestral protein emerged in an anoxic environment when the demands for aqueous nitrogen species exceeded the abiotic supply. The implications for martian ecosystems is that nitrogenase would have also emerged within an anaerobic subsurface environment, not in the oxic surface environment.

Experiments on the effects of low $pN_2$ on diazotrophic nitrogen-fixing soil bacteria have shown that they could grow in $N_2$ partial pressures of 5 mbar but not 1 mbar (Klingler *et al.*, 1989). This result suggests that further experiments on wild type species are required to determine whether the evolution of $pN_2$ in the martian atmosphere was a significant deterrent to the expansion of early life, especially after Mars lost most of its atmosphere. Analyses of the nitrogen budget and of nitrogen cycling from deep subsurface environments in South Africa indicate that the $pN_2$ is higher at depth than on the surface, that most of this $N_2$ originates from the rock formations through nitrogen cycling and that $N_2$ is being actively fixed in the subsurface by microbial communities (Lau *et al.*, 2016; Silver *et al.*, 2012). Given the presence of a cryosphere barrier to diffusion on Mars, the nitrogen availability and perhaps even the $pN_2$ of subsurface brines is likely to be higher there than on the martian surface.

## 4. Biosignatures Of Past Rock-Hosted Life

Examination of fossil evidence for life on Earth prior to ~2 Ga is hindered by the fact that single-cell prokaryotes typically do not produce inorganic cellular components and Archean rocks have been subjected to metamorphic conditions capable of completely erasing the organic microscopic cellular remains. For those rare low-metamorphic grade Archean rocks, molecular biosignatures such as hopanes (Eigenbrode, 2008) and their associated isotopic signatures (Williford *et al.*, 2016) can constrain ancient metabolic processes. Examining examples of fossilized subsurface ecosystems in Phanerozoic rocks, however, provides a bridge between modern day processes and contestable Archean examples (Table 1; Fig. 4).

*4.1. Subsurface Life Biosignatures in hydrothermally-altered ultramafic rocks*

Magma-poor paleocontinental margins expose large volumes of mantle peridotites to infiltration by seawater (Whitmarsh *et al.*, 2001). Along the Lower Cretaceous Iberian margin, seismic data indicates 25-100% serpentinization-driven alteration of ultramafic crust to depths of 4 km (Dean *et al.*, 2000). The upper ~1 km is most heavily altered with serpentinized rocks cross-cut by calcite-brucite assemblages, which isotopic data indicate precipitated at temperatures from 25-40°C. The contact zone is hypothesized to represent the deep plumbing of a Cretaceous "Lost-city" type hydrothermal system (Kelley *et al.*, 2005; Klein *et al.*, 2015).

Mineralized veins in the contact zone at depths of ~750 m are significantly enriched in organic carbon. Analysis for biosignatures revealed round to rod-shaped structures, ~2–200 μm in diameter, which are consistent with the morphologies of microbial colonies. Analyses of these putative fossilized cells with Raman spectroscopy revealed them to be carbon-enriched, with C–H, –CH$_2$, and –CH$_3$ functional groups. Band positions are consistent with lipids, amino acid side chains of proteins and carbohydrates, and amide I bonds in proteins. Further analysis of lipid biomarkers revealed nonisoprenoidal dialkylglycerol diether lipids of bacterial origin and acyclic glycerol dibiphytanyl glycerol tetraether lipids of archaeal origin. Thus, hydrothermal activity ~750 m beneath the seafloor at ~120 Ma sustained very abundant archaeal and bacterial microbial communities, equivalent to ~$10^9$ cells cm$^{-2}$, within fractures leaving behind morphologic fossils, organic carbon, and lipids (Klein *et al.*, 2015).

Fossilized remains of microorganisms have also been described in carbonate or serpentine veins of ~1 Ma ultramafic peridotite rocks in the Mid-Atlantic Ridge, and characterized by a combination of morphology, chemical composition, and the presence of organic matter, sometimes including specific complex amides usually characteristic of biopolymers (Ivarsson *et al.*, 2018; Ménez *et al.*, 2012). In these systems, complex organics are found in either aragonite veins (Ivarsson *et al.*, 2018) or within a poorly crystalline mix of serpentine, magnetite, and hydrogarnet, and remnant orthopyroxene with chemical enrichments in Ni, Co, Mo, and Mn (Ménez *et al.*, 2012) that are different compared to microbial preservation in basalt-hosted systems, which are instead dominated by clay minerals or Fe-oxides (see §4.2 below).

*4.2. Fracture-filling fossilized complex subsurface chasmoendolithic and cryptoendolithic communities*

Basaltic rocks cored from modern-day continental groundwater circulation sites show that cells are strongly concentrated within clay and oxides assemblages in fractures and pore spaces (e.g. (Trias *et al.*, 2017), their Supplementary Figure 15). Encrustation of biological matter by mineralization is a key means of preserving the structures over geologic time, as cataloged for rocks of multiple ages and types (Hofmann, 2008; Hofmann and Farmer, 2000). Investigations of ancient, now fossilized fracture surfaces in igneous rocks show these are zones of concentration of microbial activity, sometimes including complex communities of organisms with multiple trophic levels, preserved by mineralization.

In seamount basaltic lavas ranging in age from 48 to 83 Ma, the fossilized remains of chasmoendolithic (fracture-dwelling) subsurface microorganisms, i.e. coccoidal, filamentous or stromatolitic structures with elevated carbon concentration and organic matter such as lipids and rare chitin are preserved in mixtures of clay, Fe-oxides, and Mn-oxides and carbonate and gypsum veins (Bengtson *et al.*, 2014; Ivarsson *et al.*, 2012; Ivarsson *et al.*, 2009; Ivarsson and Holm, 2008). The micro-stromatolitic structures are interpreted as the result of Fe- and Mn-oxidizing bacteria, and based on mineral succession, appear to be the initial colonizers of sub-seafloor basalt (Bengtson *et al.*, 2014; Ivarsson *et al.*, 2015). Most filamentous and coccoidal fossils have so far been interpreted on morphological characteristics and rare chitin as fungal hyphae and yeast growth stages, respectively (Ivarsson *et al.*, 2012; Ivarsson *et al.*, 2015). This is probably not due to a dominance of fungi in sub-seafloor crust but instead due to fungi being more easily recognized than prokaryotic fossils. Micro-stromatolites and associated single-celled features with morphologies comparable to S-cycling archaea like *Pyrodictium* species suggest prokaryotic remains as well (Bengtson *et al.*, 2014; Ivarsson *et al.*, 2015). The organic micron-size coccoidal shapes occur in concentrations equivalent to ~$10^7$ cells cm$^{-2}$ on the vein walls with vein-containing tubular ichnofossils (see §4.3 below) and with saline fluid inclusions recording entrapment temperatures of ~130ºC. Documentation by synchrotron-based X-ray tomographic microscopy (SRXTM), laser confocal Raman spectroscopy, and environmental SEM suggests the 3D-filamentous structures found at 68 to 153 meters below the seafloor within the fractured basalt are the remains of a syntrophic community of chemolithoautotrophs, hyphae-forming fungi, and microstromatolitic *Frutexites* (Bengtson *et al.*, 2014; Ivarsson *et al.*, 2012; Ivarsson *et al.*, 2015).

As a second example, in continental flood basalts of Miocene age (6-17 Ma) in Oregon, secondary minerals formed within and near fractures preserve ~1 μm size coccoidal and rod-shaped microstructures and framboidal pyrite associated with kerogen (McKinley *et al.*, 2000) in an aquifer where the present day microbial communities are comprised of sulfate reducing bacteria (Baker *et al.*, 2003). Iron oxyhydroxides, smectites, zeolites, and silica within fractures and smectite veins were investigated because they contained framboidal pyrite.

As a third example, fossilized remains of biofilms have also been found preserved in 114 Ma calcite-filled fractures in granite at a depth of 200 m in the Fennoscandian shield using TEM (Pedersen *et al.*, 1997). In the same granitic rocks, coupled bacterial sulfate reduction-anaerobic $CH_4$ oxidation paleoactivity is recorded by the -125 to +36.5‰ V-PDB $\delta^{13}C$ values and diagnostic lipid biomarkers preserved in vein-filling calcite (Drake *et al.*, 2015a) that formed at temperatures <50°C and the -50 to +91‰ V-CDT $\delta^{34}S$ values in pyrite lining open fractures (Drake *et al.*, 2015b) over a depth range of 200 to 750 meters. These paleobiosignatures are consistent with the present day observations of coupled bacterial sulfate reduction-anaerobic $CH_4$ oxidation zone over similar depth ranges in fracture water from the granite though in one instance the sulfate rich zone is above the methane rich zone (Pedersen *et al.*, 2014) and in the other instance the opposite is true (Hallbeck and Pedersen, 2012).

Some organisms, known as "autoendoliths", play a more active role in the formation of rock edifices, whose precipitation can result directly from microbial metabolism and encapsulate the responsible microbial constituents (Marlow *et al.*, 2015). For example, anaerobic methanotrophs oxidize methane, increase alkalinity, and produce bicarbonate that precipitates as carbonate rock at methane seeps (Peckmann *et al.*, 2001). Metabolic activity continues from within the rock (Marlow *et al.*, 2014b), and biosignatures of the entombed organisms can persist for hundreds of millions of years (Peckmann and Thiel, 2004).

Excellent preservation of fossilized fungal mycelia have also been reported from granites of the Fennoscandian shield with diagnostic morphologies like anastomosis between branches (Ivarsson *et al.*, 2013). Putative fossil fungi have been reported from mineralized veins associated with the 458 Ma Lockne impact (Ivarsson *et al.*, 2013). Putative fossils of filamentous microorganisms have been reported from the 89 Ma Dellen impact-induced fractures in granites. Both sites are related to the subsequent hydrothermally formed mineralization indicating that impact-generated habitats in igneous rock are favorable for microbial colonization and preservation.

Other examples of filamentous fabrics include those preserved in chalcedony and/or zeolite in tens of terrestrial volcanic rocks ranging in age from Tertiary to Mesoproterozoic (Hofmann and Farmer, 2000); filamentous fabrics of what are interpreted as Fe-oxidizing chemotrophic bacteria of Devonian age in calcite veins cross-cutting lacustrine sedimentary rock (Trewin and Knoll, 1999); complex mineralized filamentous structures in Devonian age pillow basalt (Eickmann *et al.*, 2009; Peckmann *et al.*, 2007); and fungi-like mycelial fossils in vesicles and fractures of 2.4 Ga basalt in South Africa (Bengtson *et al.*, 2017). The interpretations that these fossils represent subsurface prokaryotes and fungi are consistent with observations of the present day subsurface biosphere (see Section 3.1). Nonetheless, as the record is pushed backward, Earth's overprinting processes demand more sophisticated high-resolution analyses for biogenecity determination (see also §4.6).

*4.3. Microbial trace fossils in recent and ancient glass*

Complex, tubular structures are sometimes found emanating from alteration mineral-filled fractures in basalts, basaltic glass, or impact glass may in some cases be trace fossils of microbial origin, called ichnofossils. Determining biotic vs. abiotic origin requires careful observations of textural subtleties and paired morphological-chemical criteria (e.g., (McLoughlin and Grosch, 2015). As with many exothermic inorganic chemical processes that are exploited by chemolithoautotrophic microorganisms distinguishing structure formed by abiotic reactions from biologically mediated ichnofossils is challenging (Grosch and McLoughlin, 2014; Knowles *et al.*, 2012; Wacey *et al.*, 2017). We describe two ancient examples involving rock-hosted microbial life below.

These structures were first reported in Pleistocene volcanic glass in Iceland (Thorseth *et al.*, 1992) and have since been documented globally in young seafloor pillow basalt glass (Fisk *et al.*, 1998; Fisk and McLoughlin, 2013; Furnes *et al.*, 1996; Thorseth *et al.*, 1995) and in pillow basalt of ancient ophiolites and greenstone belts dating back 3.5 Ga (Furnes *et al.*, 2008). The mechanisms of formation of these features combine dissolution of glass with leaching of cations and formation of clay minerals, Fe and Mn silicates, and Fe and Ti oxides (Staudigel *et al.*, 2008).

Basaltic glass samples from 1.3-1.8 km depth of the Hawaii Scientific Drilling Program core, Raman, Deep UV fluorescence and 16S rRNA sequencing show microorganisms are present in clays at the dissolution boundary with the glass near microtubular structures (Fisk *et al.*, 2003). In modern seafloor basalts on the Mohns spreading ridge of the Norwegian Sea, tubular dissolution structures originate from the palagonite-glass interfaces and evidence for bacterial processing includes the characteristic rounded and elongated, microbial size, 0.5-2 µm, pores, enrichments of Mn on the rims of coccoid-shaped structures with elevated concentrations of C, N, and organic carbon with a depleted isotopic signature (Kruber *et al.*, 2008; McLoughlin *et al.*, 2011). Ichnofossils like these modern examples are found in volcanic glass of the 92 Ma Troodos ophiolite (Furnes *et al.*, 2001) where tubular dissolution structures possess 3-D spiral morphologies with organic carbon and nitrogen enriched linings (Wacey *et al.*, 2014). Careful analyses of the relationship between infilling clay minerals and organics shows that the organics formed first by microbial extracellular materials and were then mineralized by clays.

In the 14 Ma Ries impact crater microtubules related to microbial life are found in the impact-generated glass within heavily altered zones with clay minerals and Fe oxides. The carbon-bearing materials have C-H$_x$ and amide I and II absorption bands from organic materials, not observed in tubule-free areas (Sapers *et al.*, 2014). Further analysis with Raman spectroscopy showed quinoic compounds and STXM coupled with NEXAFS showed Fe redox patterns in these areas consistent with microbially mediated dissimilatory Fe reduction (Sapers *et al.*, 2015).

*4.4. Lipid biomarkers of rock-hosted life*

Neutral lipid biomarkers have been widely utilized as a biomarker of terrestrial life in sedimentary and petroleum deposits (see review by (Brocks and Summons, 2005) and their application to Archaean marine sediments has had significant impact on the understanding of the evolution of prokaryotes (e.g. (Brocks *et al.*, 1999). However, aliphatic and polycyclic lipids in the metamorphosed Archean sediments containing polyaromatic hydrocarbons were later shown to be drilling contamination (French *et al.*, 2015). Unmentioned, and unresolved, is whether some of the low concentrations of bacterial lipids and Archaeal isoprenoids found in the rock matrix could in fact have originated from extant subsurface microorganisms colonizing the rock mass.

Analyses of lipid biomarkers in modern marine sediments under anoxic conditions document how rapidly the lipid biomarkers of marine planktonic biomass, both eukaryotic and prokaryotic, are quickly replaced within the water column and the surface seafloor sediment by the bacterial lipids of the subseafloor biosphere (Schubotz *et al.*, 2009). The archaeal lipid half-lives appear to be longer on the order of hundreds of thousands of years (Xie *et al.*, 2012). A geological test for both the age and preservation of subsurface bacterial lipids has been documented in Cretaceous age marine sediments that were intruded by mafic dikes 3.4 million years ago (Table 1; Fig. 4). Profiles of the phospholipids (active bacteria) and glycolipids (extinct bacteria) both stratigraphically and as a function of distance from the intrusion indicate that the glycolipids result from the decay of phospholipids of subsurface bacteria. These lipids from subsurface bacteria predate the 3.4 Ma intrusion and postdate the 44 Ma age of burial sterilization ($T_{max}$ = 125ºC) of the marine shale (Elliott *et al.*, 1999; Ringelberg *et al.*, 1997).

*4.5. Putative Archean Subsurface versus Surface Microbial Biosignatures*

The earliest preserved microbial structures are commonly agreed to be in the 3.4-3.5 Ga rock units of the Pilbara Craton, Australia comprised of microfossils and laminated sedimentary structures consistent with stromatolites containing contentious carbonaceous biosignatures (e.g. (Brasier *et al.*, 2006; Buick *et al.*, 1981; Noffke *et al.*, 2013; Schopf, 1993; Walter *et al.*, 1980). suggest the emergence of oxygenic photosynthesis at this time, though their biogenicity has been questioned (e.g. (Buick *et al.*, 1981; Lowe, 1994; McLoughlin *et al.*, 2008).

The 3.46 Ga Apex chert either represents hydrothermal dikes with silica-mineralization containing kerogenous microfossils involved in subsurface cycling of $CH_4$ (Schopf et al., 2018) and/or organic matter formed by abiogenic mechanisms, e.g. Fischer-Tropsch-type synthesis or remobilization of other organics (Brasier et al., 2002; Brasier et al., 2005; Brasier et al., 2006; García-Ruiz et al., 2003). Other workers have suggested the -56‰ $\delta^{13}C$ V-PDB value of $CH_4$ trapped in fluid inclusions of the same veins could be the earliest evidence of methanogenesis in the subsurface (Ueno et al., 2006).

Sulfur isotope fractionation between sulfides and barite in sediments and cross-cutting veins (Shen et al., 2009) and within seafloor basalt and komatiite (Aoyama and Ueno, 2018) of the Dresser Formation suggest that sulfate reducing bacteria were also present and metabolically active at the near surface and subsurface by ~3.46 Ga. Microfossils preserved in chertifed Strelley Pool arenite and pyrite with negative $\delta^{34}S$ V-CDT values also suggest the presence of subsurface sulfate reducing baceria at 3.43 Ga (Brasier et al., 2015). Pyritic filaments preserved in the 3.24 Ga deep sea volcanogenic massive sulfide deposits within the Sulphur Springs Group provide the most convincing evidence of early life in the form of thermophilic, chemotrophic prokaryotes living in hydrothermal systems beneath the seafloor (Rasmussen, 2000).

Putative microfossils in the form of mineralized tubular features in basaltic glass from the 3.47–3.46 Ga upper Hooggenoeg Formation of the Barberton Greenstone Belt (BGB) containing isotopically light carbonate have been proposed as subsurface biosphere fossils (Banerjee et al., 2006; Furnes et al., 2004). Titanite mineralized microtubules in 3.35 Ga basaltic glass with a minimum age of 2.9 Ga have also been suggested to represent evidence of biological processing (Banerjee et al., 2007). Very negative $\delta^{34}S$ V-CDT values from pyrite within these microtubules suggest that they were formed by microbial sulfate reduction (McLoughlin et al., 2012). More recently, however, Grosch and McLoughlin (2014) disputed the biogenecity of the microtubule textures suggesting they represent contact metamorphic textures associated with post-depositional intrusions (Table 1; Fig. 3).

Even more controversial are the earliest traces of life from 3.75-3.95 Ga amphibolite grade metamorphic rocks in Greenland and Labrador in the form of "biogenic graphite" (McKeegan et al., 2007; Mojzsis et al., 1996; Tashiro et al., 2017) and a single graphite inclusion in 4.1 Ga zircons from Jack Hill Australia (Bell et al., 2015) that yield negative $\delta^{13}C$ V-PDB values comparable to those of modern life. Determining whether they represent primary organic matter versus secondary organic matter formed during later metamorphism is challenging (Papineau et al., 2011) and isotopic values do not determine whether they formed by biological fractionation or abiotic processes (Sherwood Lollar et al., 2006; van Zuilen et al., 2003), let alone whether they represent "chemofossils" of phototrophic or subsurface chemoautotrophic microbial biomass. Other evidence for Archean life is based upon textural evidence such as putative cm-scale stromatolites in 3.7 Ga metacarbonates in Greenland (Nutman et al., 2016) and tens of micron scale hematite filaments in 4.2 Ga metasedimentary rock in Quebec (Dodd et al., 2017).

In summary, the paucity of low metamorphic grade Archaean rock record hampers our ability to identify unambiguous biosignatures older than ~3 Ga, which is approximately the time frame at which Mars' broad-scale habitability began to decline. Putative morphological biogenic structures combined with C and S isotopic evidence consistent with biological material, methanogenesis and sulfate reduction have been preserved in predominantly subsurface volcanic domains, dikes and quartzites are consistent with subsurface life as those found in marine sediments are suggestive of photosynthetic, both oxygenic and anoxygenic, life.

Noteworthy is a quote from a recent review of fossils of ancient life: "Why are few cellular fossils found in rocks before 2.5 Ga? For decades, the main search image has been cyanobacteria-like assemblages as silicified algal mats and stromatolites. Have we been looking for fossils in the wrong places?" (Brasier *et al.*, 2015). In light of new insights on the magnitude of the rock-hosted biosphere on Earth, it seems clear that while we were not looking in the wrong places, the taphonomic windows and environmental settings investigated demand expansion for research on the preserved biomarkers of ancient life. The same can be said, therefore, for the >50% of the surface of Mars that is older than ~3 Ga and unmetamorphosed as representing a particularly compelling exploration frontier.

*4.7. The Footprint of Fossil Rock-Hosted Life*

The confirmation of biosignatures often takes place at the micrometer scale by a suite of integrated techniques described in the examples above. Nevertheless, the "footprint" of fossil rock-hosted life can far exceed this scale as the mineralogical, chemical, and isotopic signatures for the presence of life can often be observed in bulk rock samples, sometimes over enormous volumes ranging from kilometers to millimeters.

An example of just how large this footprint can be is the magnetic anomalies on the scale of 10's of kilometers detected around oil fields by aeromagnetic surveys (Fig. 5). These anomalies are characteristic of fresh water oil reservoirs and results from the oxidation of hydrocarbon coupled to the reduction of Fe(III) in the sediment by Fe(III)-reducing bacteria producing fine-grained magnetite that record the ambient magnetic field at the time of crystallization over the geological lifespan of the oil reservoir (Liu *et al.*, 2004; Schumacher, 1996). The high-Fe groundwater zone in the Atlantic Coastal Plain stretches for hundreds of kilometers and is created by $Fe^{3+}$-reducing bacteria (Chapelle and Lovley, 1992).

Another example is the footprint left behind by the subsurface microbial processes in pore-waters in sediments that successively consume organic matter and give mineralogically, chemically and isotopically characteristic products that are much larger than the microorganisms forming them. For example, the association of pyrite with non-ferroan calcite with $\delta^{13}C$ values of approximately -20‰ V-PDB is a clear indication of the former presence of sulfate-reducing bacteria as has been shown in modern day deposits (Coleman *et al.*, 1993). The likely $\delta^{13}C$ value for each of the processes, which may vary from -20‰ to +15‰ V-PDB is summarized in figure 1 of Coleman (Coleman, 1985). Any part of a sedimentary succession might have evidence of one process only, and this results from the rate of burial of the sediment (or changes in rate), which controls in which zone the organic matter resides for the longest time (Irwin *et al.*, 1977). The characterization is sufficiently specific that contributions from different processes can be identified in a single mineral; for example, Mn reduction, Fe reduction and methanogenesis in siderite (Fe carbonate) minerals in 315 Ma sediments (Curtis *et al.*, 1986). These mineral biosignatures occur as intergranular pore-filling cements in many sediment rock types ranging from shale to coarse sandstone (Curtis, 1977). However, the most spectacular occurrences are large spherical or sub-spherical nodular concretions ranging in size from a few mm up to 2 m diameter (Thyne and Boles, 1989). Their visibility makes them excellent pathfinders for other more detailed analyses to confirm their origin and they can preserve biosignatures for more than 550 million years (Dong *et al.*, 2008).

Other examples of carbonate/oxide concretions produced by anaerobic and micro-aerophilic subsurface bacteria are found in ancient sandstones at scales of up to meters (Abdel-Wahab and McBride, 2001; Coleman, 1993; Mcbride *et al.*, 2003). These sandstone concretions form as a result of microbially mediated redox reactions occurring during fluid flow long after deposition of the sediment. Meter size Fe(II)-rich carbonate/iron oxide concretions (Fig. 4) are found in Jurassic sandstone deposits of southwest Colorado that were formed at hundreds of meters depth between 2 to 0.5 Ma as the Colorado River Basin was uplifted (Loope *et al.*, 2010; Mcbride *et al.*, 2003). Similar size ferroan calcite and siderite concretions occur in Late Paleocene/Early Eocene Wasatch Group sandstones and siderite nodule-bearing cores from the formation (Lorenz *et al.*, 1996) yielded thermophilic Fe(III)-reducing bacteria that were capable of producing prodigious quantities of siderite (Roh *et al.*, 2002). In subaqueous systems unconstrained by rock matrix, authigenic carbonate mounds at $CH_4$ seeps, formed from carbon mobilized by methane- and alkane-oxidizing microorganisms (Formolo *et al.*, 2004; Greinert *et al.*, 2001; Ussler III and Paull, 2008), can be hundreds of meters tall and more than a kilometer wide (Klaucke *et al.*, 2008).

At smaller scales but still larger than individual microorganisms, filamentous bacteria often form mats centimeters in size that later are silicified into stalactite-like cavity filling textures (Hofmann and Farmer, 2000). Smaller still are framboidal pyrites generated by sulfate-reducing bacteria in sizes ranging up to 10 of μm's in diameter (Maclean *et al.*, 2008; Popa *et al.*, 2004; Fig. 5). Framboidal pyrites of similar size and texture are seen in Archean sedimentary deposits (e.g. (Guy *et al.*, 2012). However, framboidal pyrite alone is not an infallible biosignature as pyrite with similar microcrystalline textures can be produced abiotically in the laboratory if the solutions are extremely supersaturated with respect to pyrite and/or the temperatures are greater than 60°C (Ohfuji and Rickard, 2005) and occur naturally too in ore deposits formed at 150° to 320°C (Halbach *et al.*, 1993), temperatures well above the limit of hyperthermophiles. Nevertheless, true framboidal pyrites are widely associated with microbial sulfate reduction and can act as a valuable pathfinder so that other characterizations can be performed.

Reduction spheroids were formerly believed to be created by detrital organic matter abiotically reducing Fe(III) minerals to Fe(II) minerals in red beds, but additional mechanisms, such as radiolysis and subsurface bacteria have been advanced to explain their occurrence (Hofmann, 1990; Hofmann, 1991; Hofmann, 2008; Keller, 1929). Reduction spheroids that range in diameter from millimeters to decimeters with a core enriched in uranium and vanadium have been found throughout the geological record reaching back to the oldest red beds in the Mesoproterozoic and are believed to record the activity of subsurface microorganisms often pre-dating peak metamorphism and deformation (Spinks *et al.*, 2010).

Thus, the footprint of subsurface metabolic activity can greatly exceed the organic content of the microorganisms responsible. To quantify some of the examples above, 1 to 20 cells occurring within a cluster or patch in the rock would be comprised of $2 \times 10^{-14}$ to $4 \times 10^{-13}$ grams of organic carbon in a $10^{-11}$ cm$^3$ volume. Microbially generated framboidal pyrite ranges in size and mass up to $10^{-9}$ cm$^3$ of 50 wt% S. Reduction spheroids can form $10^3$ cm$^3$ volumes of enriched Fe(II). Carbonate concretions attain volumes of $10^6$ cm$^3$ enriched in inorganic carbon. These volumes thus represent the metabolic conversion of $\sim 1 \times 10^{-9}$ to $2 \times 10^6$ moles of oxidants by reduced equivalents over time scales governed by diffusion transport of dissolved oxidants/reductants of 0.1 seconds to millions of years (Abdel-Wahab and McBride, 2001; Mcbride *et al.*, 2003; Thyne and Boles, 1989).

## 5. An *Exploration Strategy For Past Rock Hosted Life Biosignatures On Mars*

### 5.1. Lessons from Earth

The types of biosignatures of rock-hosted life include morphologic structures, organic carbon, spatial patterns in geochemistry, gases, minerals, and their isotopic signatures. When available in concert they can distinguish rock-hosted life from abiotic footprints (e.g. (McLoughlin *et al.*, 2011). The criteria for recognizing such life are not fundamentally different from those articulated for more "classic" near-surface sedimentary environments (Summons *et al.*, 2011). As summarized by (Grosch *et al.*, 2014) textural, chemical, and isotopic information (about both reservoir composition and fractionation patterns) is required, initially at sub-millimeter scale and then micrometer-scale with NanoSims, FIB-TEM, and X-ray synchrotron based studies. The nature of the rock types that warrant investigation for fossil rock-hosted life and the methods for finding and then characterizing the most promising samples are, however, different. Crystalline igneous rocks altered by groundwater and impact-altered rocks are of high priority in the search for rock-hosted life on Mars. Several heuristic principles can be extracted, based on the terrestrial experience in finding and characterizing biosignatures of rock-hosted life, discussed in §3-4. These scale down from landscape-scale to the microscopic scale (Fig. 4).

First, suitable host rock formations must be identified within the environmental parameters that support life. These include zones with a suitable temperature range during water-rock interaction (<~120°C), sufficient permeability for fluid flow or porosity for diffusive transport (can be highly heterogeneous) combined with redox couples that yield sufficient energy to provide adequate power for sustaining significant biomass concentrations. Many formations may be suitable. In particular, even rocks identified with higher temperature water-rock alteration are of interest because there will exist some contact zone or gradient where the higher temperature waters cool towards Mars ambient thermal gradients. Rocks should not have excessive overprinting by later chemical or thermal processes, which might obfuscate or destroy the interpretation of the origin of the rock-hosted life biosignatures. However, low-grade hydrothermal or metamorphic overprinting of the primary rock mineralogy is not a showstopper because such environments have yielded subsurface biosignatures on Earth, although the duration of such modifications is an important consideration..

Second, finding specific locales to search for biosignatures relies on seeking interfaces at a variety of spatial scales. Studies of terrestrial rock-hosted life have revealed there are two types of interfaces conducive to rock-hosted life: zones with redox disequilibria gradients or high permeability zones of fluid flow versus low permeability zones. The former provide energy for life and the latter ensure sufficient delivery of new material to support metabolism and removal of waste products. Fault zones, fractured rock, connected vesicles and voids, and alteration zones are locations where rock-hosted life, present and past, is detected on Earth. Meters-scale and cm-scale analyses can identify these key interfaces. Detection of the biosignatures themselves relies on smaller spatial scale (sub-millimeter) analyses of the patterning in morphology, chemistry, and isotopes.

Third, bulk rock organic carbon content over large spatial scales does not track as a key indicator of the richness of the microbial community. Heterogeneity along interfaces is expected, and most subsurface cell concentrations are clustered rather than diffuse. Based upon a review of terrestrial biomass distribution (see §3.3), any search for cell-like materials requires searching rock fracture surfaces for ~10 cell clusters ($>\sim 10^3$ cells/gm) occupying 100's of $\mu m^2$ areas or identifying seams with carbonaceous material where cell concentrations can reach $10^9$ cells/gm. Thus, the ability to detect 1000 cells/gm at 100 um spatial sampling may be a rule of thumb for evaluating candidate techniques for *in situ* biosignature prospecting for rock host.

Finally, a crucial lesson from the terrestrial record of fossil rock-hosted life is that the initially detected potential biosignature is more likely to be a suggestive mineralogical, chemical or isotopic composition, possibly in a particular shape or texture, rather than a direct detection of organic carbon enrichment from such life. This is because the products of life are more volumetrically significant than life itself. Phases that may be metabolic products of rock-hosted life or be by-products of metabolic reactions include sulfide, carbonate, oxides as well as gases trapped in fluid inclusions. The ability to interrogate their microscale textures, isotopic signatures, and trace element patterns all support the ability to identify possible biosignatures that later might be confirmed with still smaller scale analyses in terrestrial laboratories.

*5.2. The Exploration Strategy For Mars*

The Earth's crust has harbored deep life since at least 3.45 Ga, and the fossil remains are well preserved in rock with low metamorphic grade, which is promising for tracking the terrestrial fossil record as well as searching for fossils in similar age rocks from Mars (Cavalazzi *et al.*, 2011; Hofmann, 2008). Despite the lack of plate tectonics on Mars, rocks preserving ancient subsurface habitats are likely to be accessible to surface exploration without necessitating large complicated drilling operations. Crustal loading by emplacement of large volcanic edifices at Tharsis and unloading around large impact basins around Isidis and Argyre have created extensional faulting surrounding these landforms that expose thick strata of crust that once hosted groundwater flow (Edwards and Ehlmann, 2015; Ehlmann *et al.*, 2011). Active erosional processes have exposed 100's of meters of these stratigraphies in a form accessible to rovers. Examples include mineralized ridges at cm- to km-length-scale, which were conduits of groundwater flow (Quinn and Ehlmann, 2018; Saper and Mustard, 2013; Siebach and Grotzinger, 2014; Thollot *et al.*, 2012). Impact craters also provide direct exposure of subsurface material by their walls, ejecta, and uplift of materials in the central peak (Cockell and Barlow, 2002; Cockell *et al.*, 2012). The complicating factors affecting preservation of rock-hosted life biosignatures on Earth such as organic matter degradation by modern organisms and overprinting of chemical/mineralogical/isotopic signatures by metamorphic fluids are likely absent or reduced in the near subsurface of Mars. Rock-hosted life biosignatures sealed in reduced mineral phases in martian subsurface may also be less susceptible to secondary oxidation during uplift and exposure to surface oxidation than surficial porous sediments comprised of oxidized mineral phases.

The exploration strategy for searching for evidence of martian rock-hosted life parallels that employed on Earth, but with the need to progressively narrow the spatial scale of the exploration zone to efficiently target, access, and explore the best sites, given the prevalence of orbital data and – relative to the terrestrial situation— the paucity of opportunities for data collection from landed missions (Table 2). Additionally, many biosignatures are (at present) only detected with advanced laboratory analyses necessitating parsimonious sample selection coupled with acquisition of contextual data for return of those samples with promising preservation of biosignatures for rock-hosted life.

The scaled exploration strategy for rock-hosted life relies on seeking interfaces and boundaries (Table 2). Redox interfaces, indicated by mineralogy with contrasting oxidation states, can be manifest at a range of spatial scales indicating the potential for past thermodynamic disequilibria that drive metabolism. Lithological interfaces that indicate zones of focused fluid flow – fault zones, dikes, fracture networks, and connected vesicles – also are required for exchange of materials with the environment. Because of the importance of subsurface hydrology in establishing and maintaining habitable conditions, reconstructing fluid flow regimes through martian aquifers is a key priority. Volcanic rocks inherit large-scale fractures during cooling and sediment compaction and closing of pore space is less pronounced due to lower gravities. Meteorite impacts represent one of many reliable modes of fracturing rock and creating reactive surface area and permeability enhancement – as demonstrated at the Haughton and Chesapeake Bay Impact Structures (Pontefract *et al.*, 2016)– all of which improve habitability prospects (Cockell *et al.*, 2012).

As an example of the exploration strategy for Mars, orbit-based data can identify rocks' lithologies recording the paleoenvironmental conditions with groundwater flowing through sulfate- and serpentine-containing rocks at Northeast Syrtis Major. The presence of serpentine alongside oxidized sulfur indicates the possible presence of redox interfaces. Abundant fracturing in the area and presence of secondary minerals suggest lithological interfaces and substantial fluid flow. Calculations of Gibbs free energy suggest that these martian habitats had the necessary energy to support anaerobic oxidation of $CH_4$ (Marlow *et al.*, 2014a). Mobile surface explorers with camera and instruments for remote assessment of mineralogy and chemistry can then pinpoint lithologic and mineralogical interfaces such as fractures, redox fronts, and zones of low temperature aqueous mineralization: for example, cross cutting serpentine veins, serpentine-carbonate contacts, and zones of intense magnetite precipitation to meters then centimeters in scale using remote sensing packages. Advanced instruments for petrology employed in contact with the rock then examine a variety of initial observables, characteristic of sites hosting signatures of paleo rock-hosted life, including organics content-mineral associations (Table 1; Table 2). In select cases (e.g. complex mineralized filaments, reduction spheroid, concretion or framboidal pyrite), a biosignature may be deemed highly likely, especially if concentrations of organic matter are associated with them. However, the best confirmation of biogenecity would require further higher resolution laboratory analyses on Earth that include significant sample preparation and nanometer scale analyses.

## 6. Summary and Recommendations for Future Directions

A review of the published studies on abundance and diversity of extant terrestrial subsurface life, the diverse environments in which it is found, their fossil remains and biomarkers and a comparison of the evolution of key metabolic pathways for phototrophic versus chemolithoautotrophic microorganisms provides guidance to the search for biomarkers of subsurface life on Mars. First, the metabolic pathways for microorganisms found in the terrestrial subsurface evolved much earlier in Earth's history than those of surface dwelling phototrophic microorganisms. Second, time-equivalent environments on Mars were much less stable than on Earth and the surface environments were challenged by radiation, aridity, freezing temperatures, and frequent obliquity-driven climate change that reduced the availability of water.

Subsurface environments inhabited by rock-hosted life are common, not rare, on Earth. Rock-hosted life is found in ultramafic serpentinizing systems, deep groundwater systems, hydrothermal systems, and shallow aquifer and diagenetic environments. Terrestrial subsurface biomass concentration tends to be highest at chemical redox gradients and at permeability interfaces; it does not correlate directly with the abundance of organic carbon. Rock-hosted life does not rely upon metabolizing organic photosynthate supplied by Earth's phototrophic organisms but rather upon the flux of inorganic energy substrates (water-rock chemical reactions, radiolysis) and the abiotic and biotic recycling of carbon and metabolic waste products. The terrestrial rock record reveals examples of subsurface biomarkers at least back hundreds of millions of years and likely to 3.45 Ga. Several excellent examples of rock-hosted life with high-quality preservation are found in rocks quite different from those traditionally explored for the photospheric-supported biosphere.

These findings suggest a well-defined exploration strategy for rock-hosted life on Mars:
(1) locate rocks preserving aquifers, i.e., the plumbing of hydrothermal and groundwater systems;

(2) then, identify redox interfaces and permeability/porosity boundaries preserved within the rock outcrop;
(3) search for locations where these interfaces exhibit mineralization that may have entombed cells;
(4) at submillimeter-scale, interrogate these zones of mineralization for patterns in organic molecule concentration, morphologies suggestive of microbial filaments or cells, changes in isotopic signatures (particularly of C, N, S and Fe), and associations between these putative biosignatures.

Armed with this strategy, evidence of the biosignatures of rock-hosted life can be found *in situ* on Mars and the best samples identified for return to Earth and further interrogation to understand the history of rock-hosted life on Mars.

**Acknowledgement**

This work has greatly benefited from the discussions and intellectual contributions of the 20 participants in the Rock Hosted Life Workshop February 8-10, 2017 at Caltech as well as the participants in the 4 pre-workshop telecons open to the community. Thanks to Mary Voytek and Michael Meyer at NASA Headquarters for workshop funding and to Penny Boston and the NASA Ames Research Center meeting support team for providing web-hosting for the telecons and their recording; all the information can be found at:
http://web.gps.caltech.edu/~rocklife2017/. We thank partial support of TCO through a subcontract supported by NASA Exobiology Award NASA NNX17AK87G to Andrew Schuerger of the University of Florida. The contribution of MC was carried out partly at the Jet Propulsion Laboratory (JPL), California Institute of Technology, under contract with the National Aeronautics and Space Administration (NASA), and via Grant NNA13AA94A issued through the NASA Science Mission Directorate and supported by the NASA Astrobiology Institute.

**Author Disclosure Statement**
No competing financial interests exist.

Peckmann, J., Bach, W., Behrens, K., and Reitner, J. (2007) Putative cryptoendolithic life in Devonian pillow basalt, Rheinisches Schiefergebirge, Germany. *Geobiology*, 6, 125–135.

Peckmann, J., Reimer, A., Luth, U., Luth, C., Hansen, B.T., Heinicke, C., Hoefs, J., and Reitner, J. (2001) Methane-derived carbonates and authigenic pyrite from the northwestern Black Sea. *Marine Geology*, 177, 129-150.

Peckmann, J., and Thiel, V. (2004) Carbon cycling at ancient methane–seeps. *Chemical Geology*, 205, 443-467.

Pedersen, K., Bengtsson, A.F., Edlund, J.S., and Eriksson, L.C. (2014) Sulphate-controlled Diversity of Subterranean Microbial Communities over Depth in Deep Groundwater with Opposing Gradients of Sulphate and Methane. *Geomicrobiology Journal*, 31, 617–631.

Pedersen, K., Ekendahl, S., Tullborg, E.-L., Furnes, H., Thorseth, I., and Tumyr, O. (1997) Evidence of ancient life at 207 m depth in a granitic aquifer. *Geology*, 25, 827–830.

Phelps, T.J., Murphy, E.M., Pfiffner, S.M., and White, D.C. (1994) Comparison between geochemical and biological estimates of subsurface microbial activities. *Microbial Ecology*, 28, 335-349.

Pinti, D.L., Mineau, R., and Clement, V. (2009) Hydrothermal alteration and microfossil artefacts of the 3,465-million-year-old Apex chert. *Nature Geoscience*, 2, 640-643.

Pontefract, A., Osinski, G.R., Cockell, C.S., Southam, G., McCausland, P.J.A., Umoh, J., and Holdsworth, D.W. (2016) Microbial Diversity of Impact-Generated Habitats. *Astrobiology*, 16, 775–786.

Popa, R., Kinkle, B.K., and Badescu, A. (2004) Pyrite framboids as biomarkers for Iron-Sulfur systems. *Geomicrobiology Journal*, 21, 193–206.

Price, A., Pearson, V.K., Schwenzer, S.P., Miot, J., and Olsson-Francis, K. (2018) Nitrate-dependent iron oxidation: A potential Mars metabolism. *Frontiers in Microbiology*, 9, 513.

Purkamo, L., Bomberg, M., Nyyssönen, M., Kukkonen, I., Ahonen, L., and Itävaara, M. (2015) Heterotrophic communities supplied by ancient organic carbon predominate in deep Fennoscandian bedrock fluids. *Microbial Ecology*, 69, 319–332.

Quinn, D., and Ehlmann, B. (2018) The deposition and alteration history of the Northeast Syrtis layered sulfates. *Journal of Geophysical Research Planets*, in revision.

Rampe, E.B., Ming, D.W., Blake, D.F., Bristow, T.F., Chipera, S.J., Grotzinger, J.P., Morris, R.V., Morrison, S.M., Vaniman, D.T., Yen, A.S., Achilles, C.N., Craig, P.I., DesMarais, D.J., Downs, R.T., Farmer, J.D., Fendrich, K.V., Gellert, R., Hazen, R.M., Kah, L.C., Morookian, J.M., Peretyazhko, T.S., Sarrazin, P., A.H.Treiman, Berger, J.A., Eigenbrode, J., Fairén, A.G., Forni, O., Gupta, S., Hurowitz, J.A., Lanza, N.L., Schmidt, M.E., Siebach, K., Sutter, B., and Thompson, L.M. (2017) Mineralogy of an ancient lacustrine mudstone succession from the Murray formation, Gale Crater, Mars. *Earth and Planetary Science Letters*, 471, 172–185.

Rapin, W., Chauviré, B., Gabriel, T.S.J., McAdam, A.C., Ehlmann, B.L., and Hardgrove, C. (2018) In situ analysis of opal in Gale crater, Mars. *Journal of Geophysical Research: Planets*, 123, https://doi.org/10.1029/2017JE005483.

Rasmussen, B. (2000) Filamentous microfossils in a 3,235-million-year-old volcanogenic massive sulphide deposit. *Nature*, 405, 676-679.

Rempfert, K.R., Miller, H.M., Bompard, N., Nothaft, D., Matter, J.M., Kelemen, P., Fierer, N., and Templeton, A.S. (2017) Geological and geochemical controls on subsurface microbial life in the Samail Ophiolite, Oman. *Frontiers in Microbiology*, 8, 56.
48

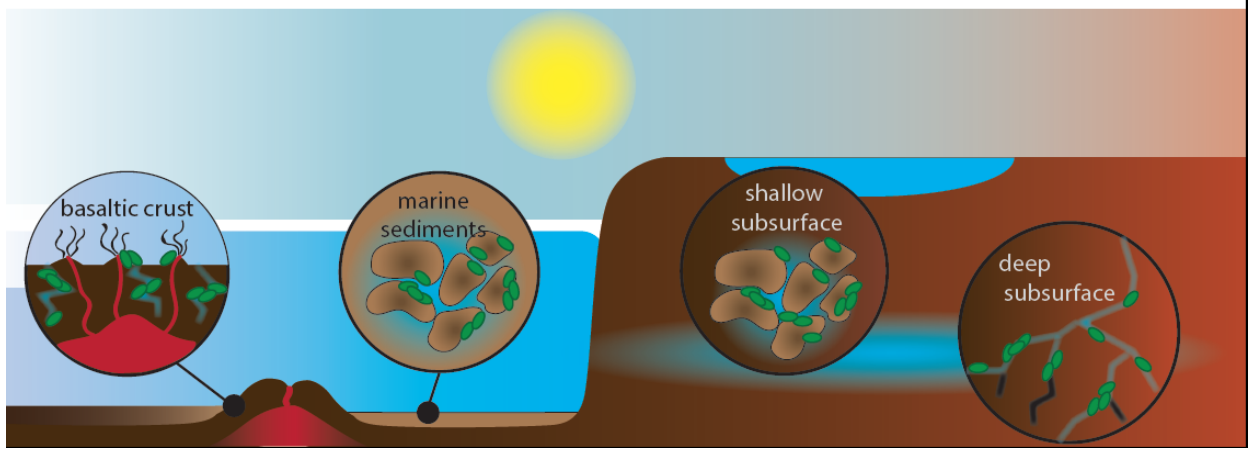

**FIG. 1.** Subsurface biosphere habitats for the Earth from left to right: Basaltic crust – $H_2$ chemolithotrophic communities fueled by advective fluid interacting with mafic rock along with abiotic hydrocarbons oxidized to carbonate mounds; Marine Sediments – primarily heterotrophic communities in a high porosity environment with diffusive flux fueled by organic photosynthate in some places and chemolithotrophic oxidation in others; Continental sedimentary aquifers of lower porosity than marine sediments and mixed heterotrophic and chemolithotrophic communities; and Deep Subsurface – fractured rock aquifers in mafic and siliceous igneous and metamorphic rock in some cases saturated with water or ice and in other cases unsaturated deep vadose zone or gas-rich reservoirs.

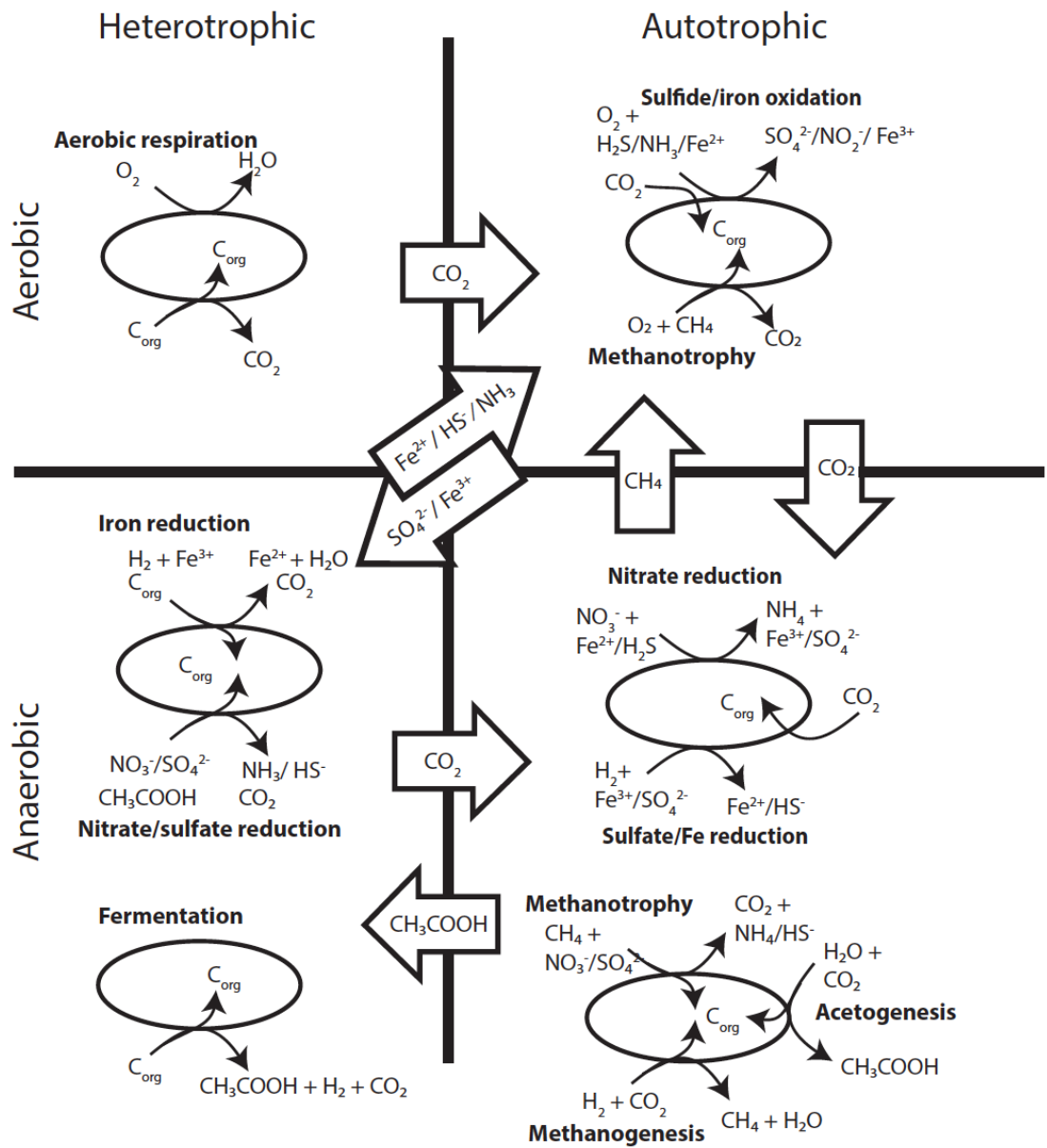

**FIG. 2.** Cartoon of different microbial metabolic processes separated into Aerobic (top), Anaerobic (bottom), Heterotrophic (left) and Autotrophic (right) bins.

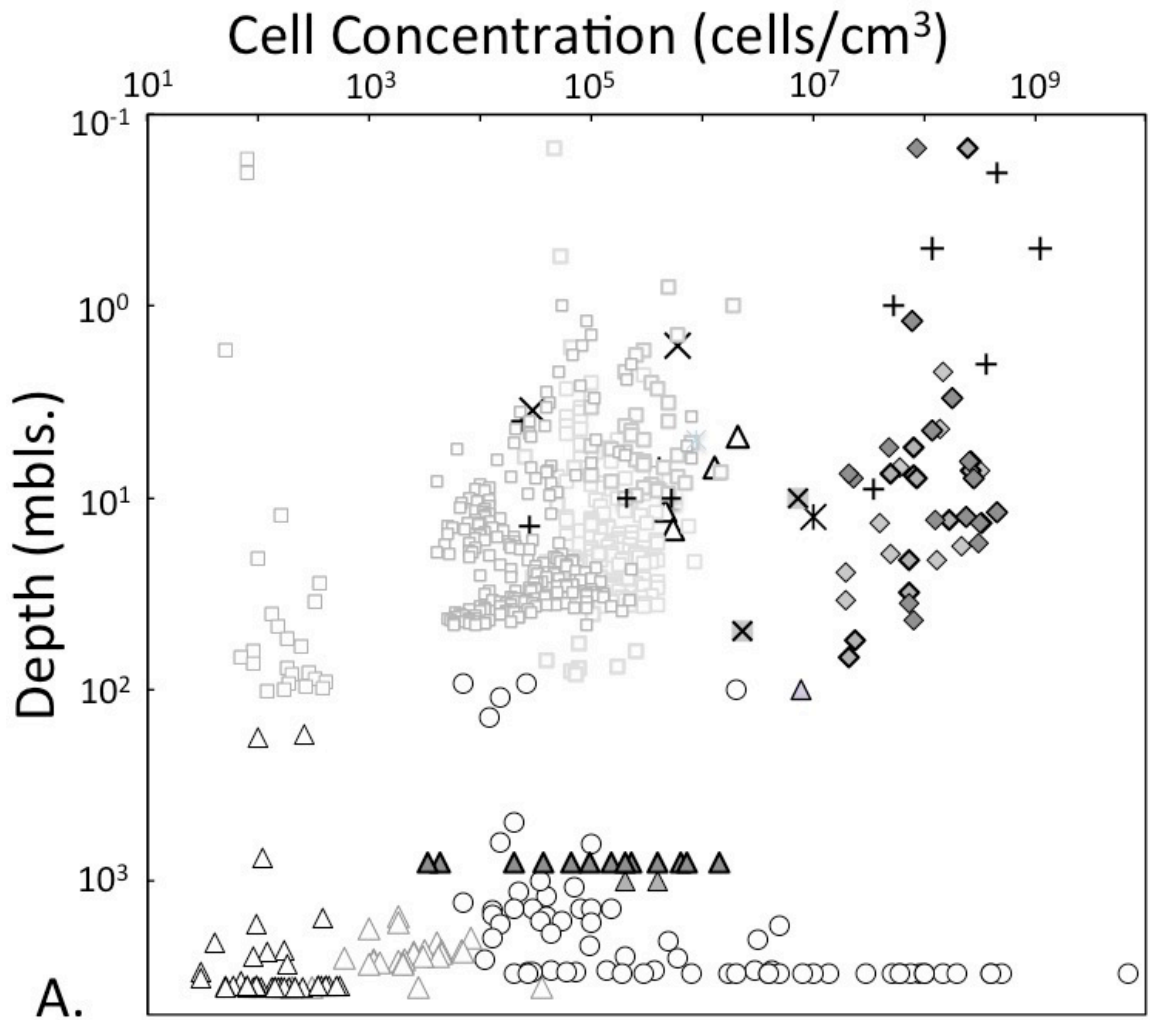

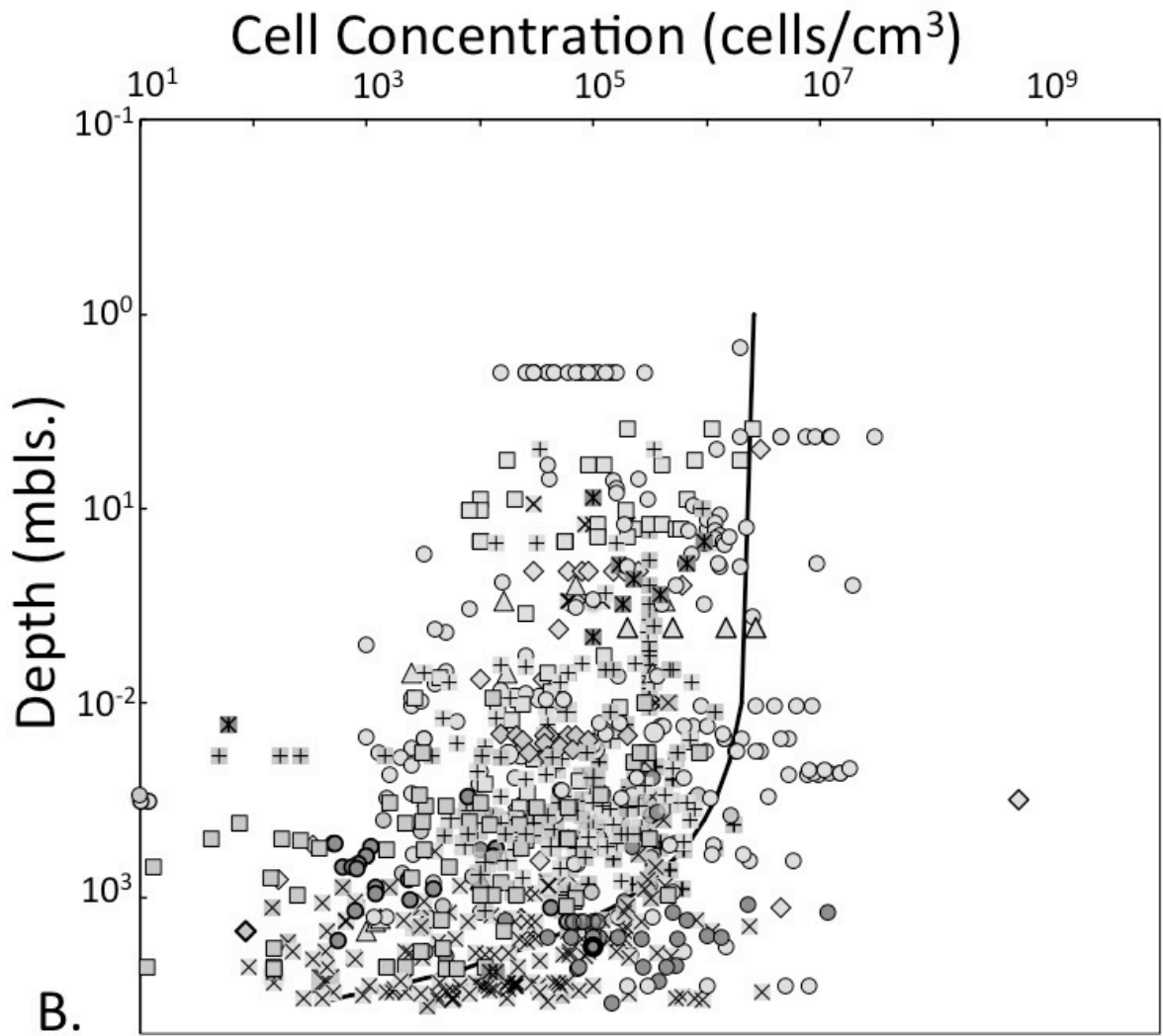

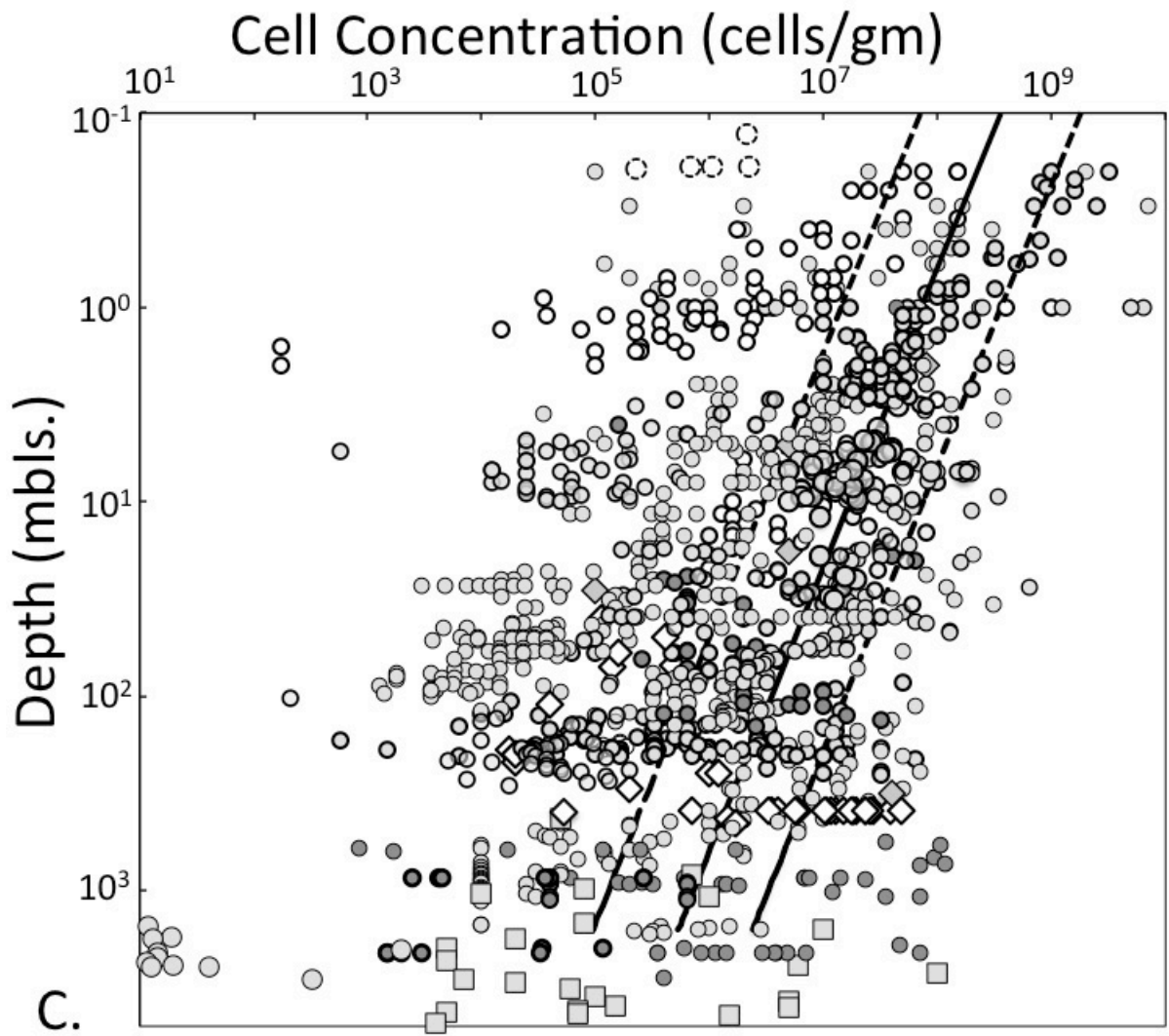

**FIG. 3.** A. Cell concentrations versus depth for ice sheets, sub-glacial sediment and permafrost. Squares = Tibetan ice sheets, gray diamonds = Siberian permafrost, asterisk = Siberian cryopeg; open triangles = Antarctica ice sheets, gray triangles = Antarctic sub-glacial sediment, cross = Antarctic permafrost; gray filled cross = sub-glacial sediment in New Zealand, plus sign = Canadian High Arctic and Svalbard permafrost, and open circles = Greenland ice sheet. Data are from Magnabosco *et al*. (2018a). B. Suspended cell concentrations in groundwater versus depth. Gray diamonds = volcanic aquifers, gray square-plus = granite aquifers, gray square = mixed metamorphic rock types, gray square asterisk = ultramafic/mafic intrusive rock aquifer, gray square cross = metasedimentary rock aquifer, grey triangle = limestone/marble aquifer, light gray circle = sedimentary rock aquifer and dark gray circle = sedimentary rock hosting oil, gas or coal. Data are from Magnabosco *et al*. (2018a) Solid black line is proposed model by McMahon and Parnel (McMahon and Parnell, 2014) for suspended cell concentration versus depth in groundwater. C. Cell concentrations versus depth for rock and soil cores. Light gray open circle = water saturated sediments or sedimentary rock, open circle = vadose zone sediments or sedimentary rock, dark gray circle = oil-gas-coal bearing sediment or sedimentary rock, gray diamond = water saturated volcanic rock, open diamond = vadose zone volcanic rock, gray square = metamorphic rock. Data are from Magnabosco *et al*. (2018a) dashed open circles = Atacama desert soil from Connon *et al*. (Connon *et al.*, 2007) and Lester *et al.*(Lester *et al.*, 2007) Solid and dashed lines represent the best-fit power law for sub-seafloor sediments proposed by Parkes *et al.*(Parkes *et al.*, 2014)

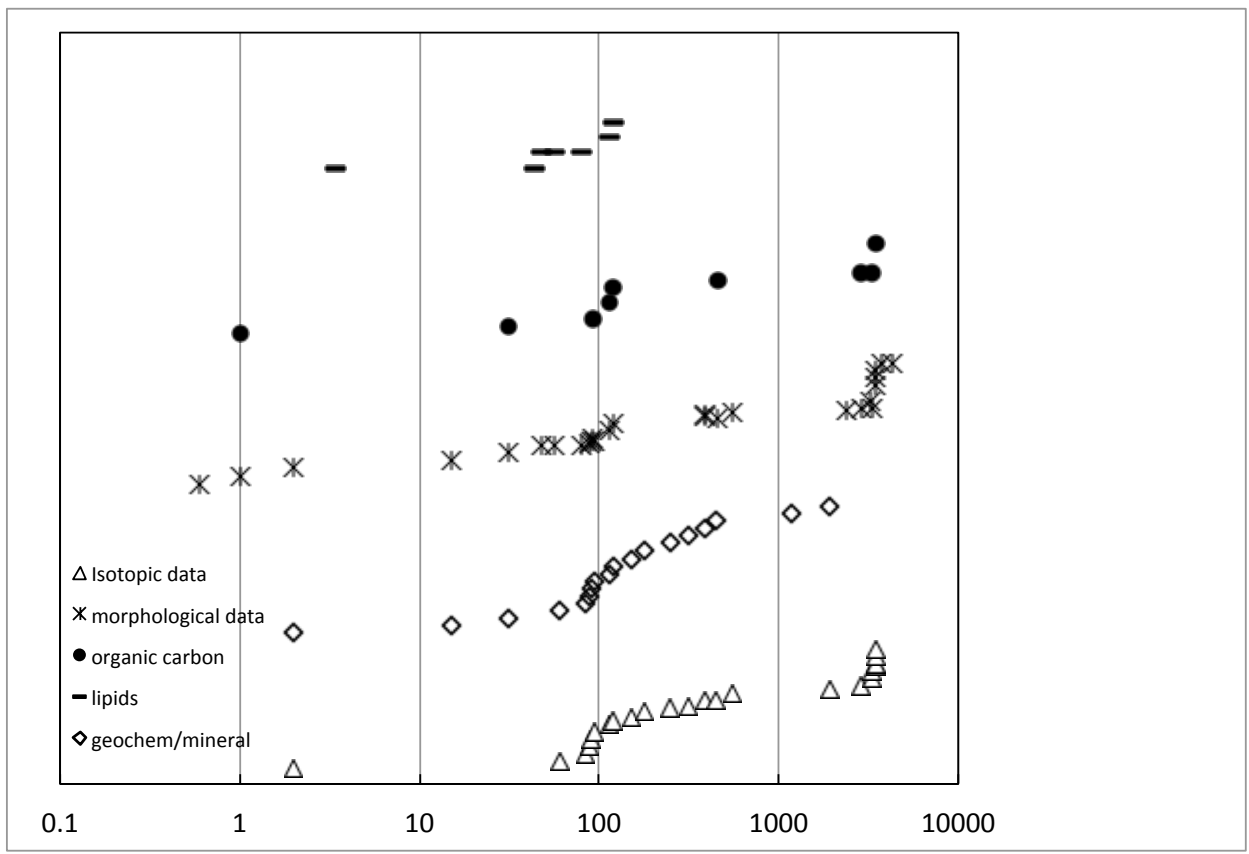

**FIG. 4.** Biomarker type versus geological age in millions of years.

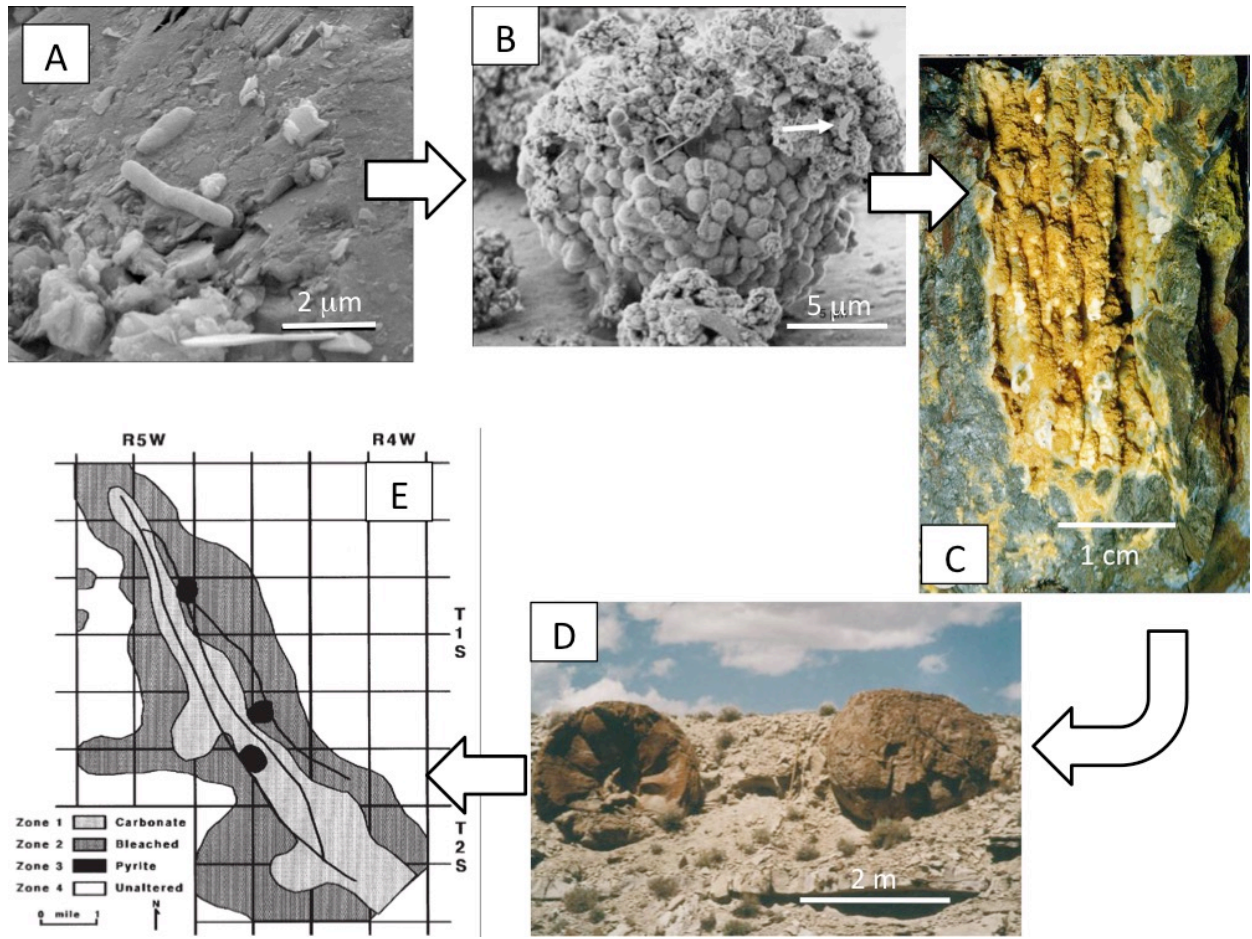

**Fig. 5.** Increasing scale of metabolic footprint. A. Single microbial cells attached to clay minerals of a 2.8 km deep fracture zone, 2 μm scale bar (Wanger *et al.*, 2006). B. Framboidal pyrite sack from 1.5 km deep borehole, 5 μm scale bar. White arrow points to single bacterial cell (Maclean *et al.*, 2008). C. "Pseudostalactite" of quartz and goethite cemented filaments occurring in Tertiary volcanic rocks in California (Hofmann and Farmer, 2000). D. Ferroan carbonate septarian concretions from 88.5 Ma the Ferron Sandstone Member of the Mancos Shale Formation in Utah that are 1 to 4 meters in diameter, 2 m scale bar (Mcbride *et al.*, 2003). E. Surface diagenetic alteration zones and traces of pre-Permian faults over Velma field, Stephens County, Oklahoma, 1 mile scale bar (Al-Shaieb *et al.*, 1994).

Table 1. Subsurface microbial biomarkers preserved in geological record

| Age (Ma) | isotope | geochemistry/ mineralogy | morphology | organic carbon | lipid | forms | Formation | ref. |
|---|---|---|---|---|---|---|---|---|
| 0.6 | | | x | | | filaments | Sea Mounts | 1 |
| 1 | | | x | x | | filaments | Ries Impact Crater | 2, 3 |
| 2 | x | x | x | | | concretions | Navajo Sandstone | 4 |
| 3.4 -44 | | | | | x | | Cretaceous shale | 5, 6 |
| 15 | | x | x | | | microcolonies | Columbia River Basalt | 7 |
| 31 | | x | x | x | | ichnofossils | sea floor basalt | 8 |
| 48 | | | x | | x | filaments | Sea Mounts | 9, 10 |
| 56 | | | x | | x | filaments | Sea Mounts | 9 |
| 60 | x | x | | | | concretions | Moeraki Formation | 11 |
| 81 | | x | | | x | filaments | Sea Mounts | 9 |
| 84 | x | x | | | | concretions | Gammon Shale | 12 |
| 88.5 | x | x | x | | | concretions | Mancos Shale Formation | 13 |
| 91 | x | x | x | | | concretions | Frontier Formation | 13 |
| 95 | x | x | x | | | concretions | Frontier Formation | 13 |
| 92 | | | x | x | | ichnofossils | Troodos ophiolite | 14, 15 |
| 114 | x | x | x | x | x | microcolonies | Fennoscandian shield granite | 16-18 |
| 120 | x | x | x | x | x | microcolonies | Southern Iberia Abyssal Plain | 19 |
| 152 | x | x | | | | concretions | Kimmeridge Clay | 20 |
| 180 | x | x | | | | concretions | Upper Lias | 21 |
| 250 | x | x | | | | reduction spheroids | Mercia Mudstone Group | 22 |
| 315 | x | x | | | | concretions | Lower Westphalian coal | 23 |
| 385 | x | x | x | | | filaments | Arnstein pillow basalt | 24, 25 |
| 388 | | | x | | | filaments | Tynet Burn limestone | 26 |
| 443 | x | x | | | | ichnofossils | Caledonian ophiolite | 27 |
| 458 | | | x | x | | filaments | Lockne Impact structure | 28 |
| 551 | x | | x | | | concretions | Doushantuo Formation | 29 |
| 1175 | | x | | | | reduction spheroids | Bay of Stoer Formation | 22 |
| 1950 | x | x | | | | ichnofossils | Jormua ophiolite complex | 30 |
| 2400 | | | x | | | filaments | Ongeluk Formation basalt | 31 |
| 2900-3350 | x | | x | x | | ichnofossils | Euro Basalt | 32, 33 |
| 3240 | | | x | | | filaments | Sulphur Springs Group | 34 |
| 3300 | x | | | | | isotope fractionation | Barberton Greenstone Belt | 35 |
| 3460 | x | | | | | isotope fractionation | Dresser Formation | 36 |
| 3465 | x | | x | x | | microcolonies | Apex Chert | 37 |
| 3465 | | | x | | | microcolonies | Apex Chert | 38 |
| 3465 | x | | | | | isotope fractionation | Apex Chert | 39 |
| 3465 | x | | x | | | ichnofossils | Hooggenoeg Formation | 40, 41 |
| 3770-4280 | | | x | | | filaments | Nuvvuagittuq belt | 42 |

[1] (Ivarsson *et al.*, 2015), [2] (Sapers *et al.*, 2014), [3] (Sapers *et al.*, 2015), [4] (Loope *et al.*, 2010), [5] (Ringelberg *et al.*, 1997), [6] (Elliott *et al.*, 1999), [7] (McKinley *et al.*, 2000), [8] (Cavalazzi *et al.*, 2011), [9] (Ivarsson *et al.*, 2009), [10] (Ivarsson *et al.*, 2012), [11] (Thyne and Boles, 1989), [12] (Coleman, 1993), [13] (Mcbride *et al.*, 2003), [14] (Furnes *et al.*, 2001), [15] (Wacey *et al.*, 2014), [16] (Pedersen *et al.*, 1997), [17] (Drake *et al.*, 2015a), [18] (Drake *et al.*, 2015b), [19] (Klein *et al.*, 2015), [20] (Irwin, 1980), [21] (Coleman and Raiswell, 1995), [22] (Spinks *et al.*, 2010), [23] (Curtis *et al.*, 1986), [24] (Eickmann *et al.*, 2009), [25] (Peckmann *et al.*, 2007), [26] (Trewin and Knoll, 1999), [27] (Furnes *et al.*, 2002), [28] (Ivarsson *et al.*, 2013), [29] (Dong *et al.*, 2008), [30] (Furnes *et al.*, 2005),

[31](Bengtson *et al.*, 2017), [32] (Banerjee *et al.*, 2007), [33] (McLoughlin *et al.*, 2012), [34] (Rasmussen, 2000), [35] (Ohmoto *et al.*, 1993), [36] (Shen and Buick, 2004), [37] (Schopf *et al.*, 2018); [38](Pinti *et al.*, 2009), [39] (Ueno *et al.*, 2006), [40] (Banerjee *et al.*, 2006), [41] (Furnes *et al.*, 2004), [42] (Dodd *et al.*, 2017)

**Table 2.** Steps to Search for Rock-Hosted Life on Mars

| Step | Spatial Scale | Key Measurement Requirements |
|---|---|---|
| 1. Identify rocks with ancient subsurface habitats | <100 m sampling | Ability to identify water-related mineral deposits from orbit and determine stratigraphic context |
| 2. Locate interfaces that represent favorable locations for rock life | Meter- to cm-scale | Ability to identify redox and permeability interfaces by identification of distinct lithologic units |
| 3. Search for mineralization from fluid flow at interfaces | Centimeter- and millimeter-scale | Ability to identify silica, carbonate, sulfate, phyllosilicate, and oxides that may mineralize microbial life |
| 4. Search for organics, mineralization, and isotopic anomalies at the interface | <100 μm sampling | Ability to detect organics, chemical, mineralogic, and/or isotopic differences between interface rocks and surrounding rocks indicative of biosignatures |
| 5. Map putative biosignatures in 3-dimensions, tracking chemical and organic variations with texture | <1 μm sampling in 3-dimensions | Ability to identify microbial textures and distinguish biotic and abiotic processes to definitively confirm fossil rock-hosted life |

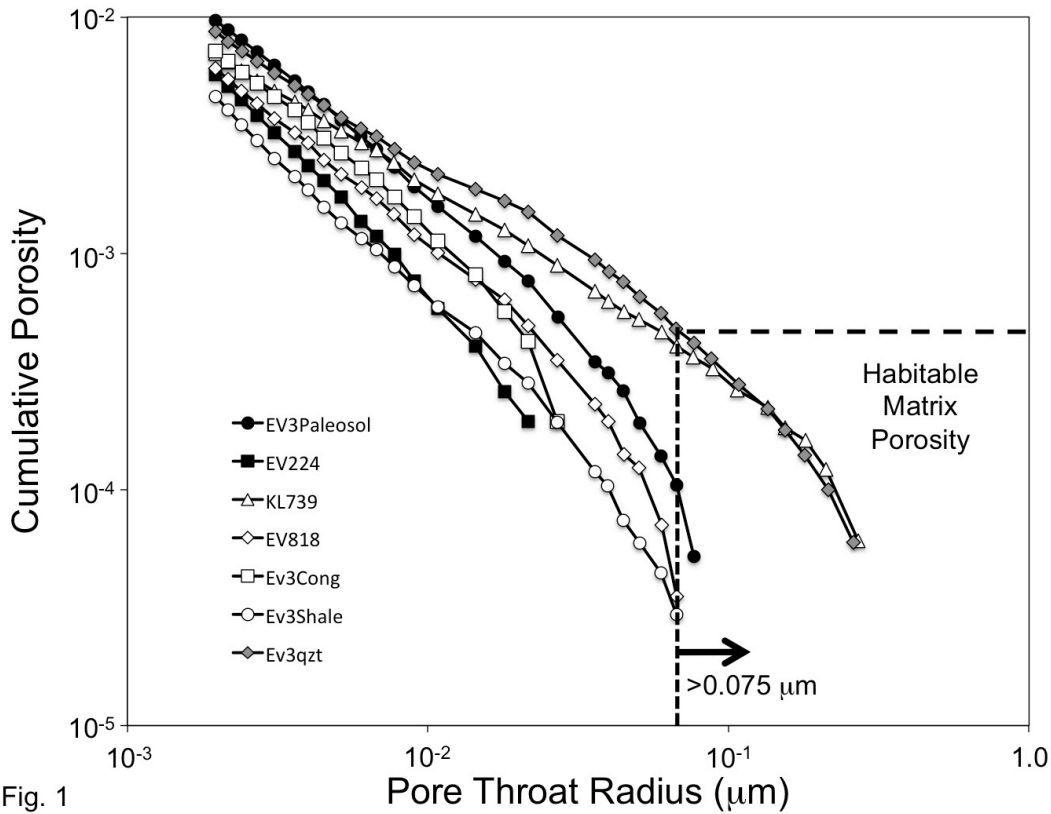

Fig. 1

**Supplementary Fig. 1.** Cumulative porosity versus maximum pore throat size based upon mercury porosimetry analyses of 2.7 to 2.9 Ga sedimentary and volcanic units from the Witwatersrand Basin of South Africa. Pore throats that are large enough to permit access by 0.15 µm diameter bacteria comprise a small fraction of the total porosity. The inaccessible porosity however does provide storage for nutrients accessible by diffusion or extracellular filaments.